\documentclass[a4paper,11pt]{article}
\pdfoutput=1
\usepackage{jheppub_kim} 
\usepackage[T1]{fontenc}
\usepackage{color}
\usepackage{float}
\usepackage{url}
\usepackage{braket}
\usepackage{amsmath}
\usepackage{amsthm}
\usepackage{amssymb}
\usepackage{amsfonts}
\usepackage{graphicx}
\usepackage{caption}
\usepackage{subcaption}
\usepackage{multirow,array}
\usepackage{qcircuit}
\usepackage{hyperref}
\usepackage{comment}
\definecolor{red}{rgb}{1,0.,0}

\usepackage{xcolor}
\colorlet{myPurple}{blue!40!red}

\usepackage{color}

\title{\boldmath Diagnosing First and Second Order Phase Transitions with Probes of Quantum Chaos  }

\author[a]{Kyoung-Bum Huh,}
\author[b]{Kazuki Ikeda,}
\author[a]{Viktor Jahnke,}
\author[a]{and Keun-Young Kim}

\affiliation[a]{School of Physics and Chemistry, Gwangju Institute of Science and Technology, Gwangju 61005, Korea}
\affiliation[b]{Department of Physics, Osaka University, Toyonaka, Osaka 5600043, Japan}

\emailAdd{hkabell1689@gist.ac.kr}
\emailAdd{kazuki7131@gmail.com}
\emailAdd{viktorjahnke@gist.ac.kr}
\emailAdd{fortoe@gist.ac.kr}

\abstract{
We explore quantum phase transitions using two probes of quantum chaos: out-of-time-order correlators (OTOCs) and the $r$-parameter obtained from the level spacing statistics.   
In particular, we address $p$-spin models associated with quantum annealing or reverse annealing. Quantum annealing triggers first-order or second-order phase transitions, which is crucial for the performance of quantum devices. 
We find that the time-averaging OTOCs for the ground state and the average $r$-parameter change behavior around the corresponding transition points, diagnosing the phase transition. Furthermore, they can identify the order (first or second) of the phase transition by their behavior at the quantum transition point, which changes abruptly (smoothly) in the case of first-order (second-order) phase transitions.
}

\begin{document}
\maketitle
\section{Introduction}

In recent years, out-of-time-order correlators (OTOCs) \cite{larkin1969quasiclassical} have gained renewed interest, especially because of their connection with many-body quantum chaos and black hole physics \cite{Kitaev14,Maldacena:2015waa}. See \cite{Jahnke:2018off}, for a recent review. For works regarding the experimental measurement of OTOCs, see for instance \cite{2019PhRvX...9a1006Y, 2017NatPh..13..781G,2017PhRvX...7c1011L,2019NatCo..10.1581L}.

An OTOC is defined as
\begin{equation}
    F_{V W}(t) = \langle W(t) V(0) W(t) V(0) \rangle\,,
\end{equation}
where $V$ and $W$ are some simple operators and the expectation value is usually taken in a thermal state. The idea is that chaotic behavior leads to an early-time exponential\footnote{{See Sec. 3.1 for subtleties and clarifications about the exponential behavior.}} suppression and the late-time vanishing of OTOCs, which happens for almost any choice of operators $V$ and $W$. On the other hand, the absence of chaos is expected to result in a non-universal, operator-dependent behavior of $F_{V W}(t)$.

Another way to characterize quantum chaotic behavior is through the statistics of the spacing between consecutive energy eigenvalues, which is usually known as {\it level spacing statistics}. In chaotic systems, the level spacing statistics obeys a Wigner-Dyson distribution \cite{BGS}, while in integrable systems it follows a Poisson distribution \cite{BTZ-Poisson,Berry-80}. We can also study the so-called $r$-parameter statistics, which is computed from the level spacing information. The average $r$-parameter takes different values depending on whether the system is chaotic or integrable. See Sec.~\ref{sec:LS} for more details.

The level spacing statistics is defined at zero temperature and involves time-scales much bigger than the ones involved in the behavior of OTOCs, which are usually computed at finite temperature. Despite these differences, both the level spacing statistics and the OTOCs can probe chaotic behavior, presumably capturing different aspects of it, due to the thermal nature of OTOCs.

Thermal effects can be removed by considering ground state OTOCs, in which the expectation value is taken in the ground state of the system. In this case, the OTOCs are completely controlled by the ground state physics \cite{Dag:2019yqu} and may become sensitive to changes in the ground state, i.e., they may diagnose quantum phase transitions \cite{Dag:2019yqu, doi:10.1002/andp.201900270,2018arXiv181201920W}.
As far as we know, the relation between level spacing statistics and quantum phase transitions remains largely unexplored\footnote{The intuition that chaos probes could also detect phase transitions appears in \cite{Garcia-Garcia:2019poj}, in the context of the mass deformed SYK model. This idea was further explored in \cite{nosaka2020quantum, nosaka2020chaos}.}. It is then interesting to explore if the tools used to characterize level spacing statistics are also useful in diagnosing phase transitions. This is one of the goals of this paper. This could help to clarify the role played by chaos in quantum phase transitions. 

Quantum phase transitions are transitions
between different phases of matter that occur at zero temperature by variation of a non-thermal control parameter. Quantum phase transitions found applications in the description of several properties of condensed matter systems \cite{Vojta_2003}, as well as in quantum computation algorithms. In this work, we will be interested in quantum phase transitions that are triggered by a method of adiabatic quantum computation known as {\it quantum annealing}.

Quantum annealing \cite{PhysRevE.58.5355} is a meta-heuristic quantum method for solving combinatorial optimization problems and could be as powerful as universal computation when a system with a non-stoquastic Hamiltonian evolves in an adiabatic way \cite{2000quant.ph..1106F}. 
The conventional form of quantum annealing is done by the quantum mechanical time evolution of an initial state, which is the ground state of a transverse magnetic field. The initial phase consists of the uniform superposition of all possible classical states, which is called the quantum paramagnetic (QP) phase. In general, the system evolves into a different phase and the performance of a quantum annealer depends heavily on the choice of transition paths. According to the adiabatic theorem, the computational time that is needed to efficiently obtain the ground state is proportional to the inverse square of the minimal energy gap $\Delta$ between the ground state and the first excited state. A lot of examples indicate that $\Delta$ decays polynomially if a phase transition is second-order \cite{Damski_2013,PhysRevB.71.224420}, whereas $\Delta$ decays exponentially if it is first-order (although there are some exceptions \cite{PFEUTY197079,PhysRevLett.109.030502,doi:10.7566/JPSJ.82.114004}). Therefore, the problem in the case of systems with second-order phase transitions is efficiently solved. There are some models for which the first-order phase transitions can be avoided when a non-stoquastic Hamiltonian or reverse annealing \cite{Perdomo-Ortiz2011} is used. The $p$-spin model we address in this article is a typical example (see also \cite{PhysRevA.95.042321,2019arXiv191107186P}). The ground state of the $p$-spin model is ferromagnetic (F) and, depending on the path, the QP/F transition becomes first-order or second-order. In \cite{2011PhRvA..83b2327F, PhysRevA.95.042321}, it is suggested that entanglement measures are good tools to analyse phase transitions of such a fully connected spin system.  


In this work, we aim at diagnosing quantum phase transitions triggered by quantum annealing using two probes of quantum chaos, namely, OTOCs and the $r$-parameter statistics. The relation between the second-order phase transition and the time average of OTOCs is reported \cite{Dag:2019yqu, doi:10.1002/andp.201900270,2018arXiv181201920W}, but not proven in general. For a better understanding, it will be useful if we can check it in other examples. In addition, less is known about the relation between the order of the phase transition and the behavior of the OTOC. As far as we know, the relation between phase transitions and the average $r$-parameter has not been studied either. In this paper, we show that OTOCs and the average $r$-parameter not only detect the phase transition but also distinguish the order of the phase transition. 
To our best knowledge, our paper is the first work in the following aspects: (1) relating both first-order and second-order phase transitions with the dynamics of OTOCs and the $r$-parameter statistics, (2) using probes of quantum chaos to study quantum annealing or adiabatic quantum computing and (3) the Hamiltonians we employed are non-stoquastic, which means the models cannot be efficiently simulated by classical ways. Some stoquastic models are studied elsewhere \cite{Heyl_2018,2017PhRvB..96e4503S,PhysRevLett.123.140602, shukla2020outoftimeorder}.  


This article is organized as follows. In Sec.~\ref{sec:mode} we review and explain the phase transition associated with quantum annealing of the $p$-spin model. In Sec~\ref{sec:PT} we use the time average of OTOCs to characterize the QP/F transition.  In Sec.~\ref{sec:LS}, we study the $r$-parameter statistics and show that the average $r$-parameter is sensitive to the phase transition. In Sec.~\ref{sec:RA} we address phase transitions associated with reverse annealing and provide more evidence that they can be diagnosed by the time average of OTOCs. Finally, we present our conclusions in Sec.~\ref{sec:fin}.

\section{Model}\label{sec:mode}
Let us start with a Hamiltonian of adiabatic quantum computation
\begin{equation} \label{H111}
    H(s)=sH_\mathrm{T}+(1-s)H_\mathrm{I},~~~~~~~s\in[0,1] \,,
\end{equation}
where $H_\mathrm{T}$ is a target Hamiltonian and $H_\mathrm{I}$ is an initial Hamiltonian. They should not commute $[H_\mathrm{I}, H_\mathrm{T}]\neq 0$. As an initial Hamiltonian we use the widely used transverse magnetic field~\cite{PhysRevE.58.5355,2000quant.ph..1106F}:
\begin{equation} \label{trans34}
    H_\mathrm{I}=-\sum_i^NX_i\,,
\end{equation}
and as a target Hamiltonian let us consider the Hamiltonian of the $p$-spin model 
\begin{equation} \label{H000}
    H_\mathrm{T}=-N\left(\frac{1}{N}\sum_{i=1}^NZ_i\right)^p\,,
\end{equation}
where $X_i$ and $Z_i$ denote, for example, 
\begin{equation} \label{xz12}
\begin{split}
    X_1 &= \sigma_x \otimes I_2 \otimes I_2 \otimes \cdots \otimes I_2\,, \\
    Z_3 &=  I_2 \otimes I_2 \otimes \sigma_z \otimes I_2 \otimes \cdots \otimes I_2\,,
    \end{split}
\end{equation}
with $\sigma_x = \scriptsize
\left( \begin{array}{cc}0 & 1 \\ 1 & 0 \end{array} \right), \ \sigma_z = \left( \begin{array}{cc} 1 & 0 \\ 0 & -1 \end{array} \right)\,.$ 

The Hamiltonian \eqref{H111} is stoquastic.  We say a Hamiltonian is stoquastic if all off-diagonal matrix elements in the standard basis are real and non-positive, otherwise it is non-stoquastic \cite{2006quant.ph..6140B}. 
We add a term that introduces non-stoquastic antiferromagnetic(AF) interactions
\begin{equation}
    H_{\text{AF}}=+N\left(\frac{1}{N}\sum_i^NX_i\right)^2\,,
\end{equation}
in such a way that 
\begin{equation} \label{eq-hamQA}
    H(s,\lambda)=s(\lambda H_\mathrm{T}+(1-\lambda)H_\text{AF})+(1-s)H_\mathrm{I}\,.   
\end{equation}

Even though a non-stoquastic term is in general hard to simulate by quantum Monte Carlo due to the negative sign problem, sometimes it makes problems efficiently solvable by adiabatic quantum computation or quantum annealing \cite{2012PhRvE..85e1112S}. Furthermore, it is believed that adding a stoquastic Hamiltonian to \eqref{H111} is not helpful for quantum speedup.

The initial Hamiltonian ($s=0$) is $H(0,\lambda)=H_\mathrm{I}$ with any $\lambda$. The final Hamiltonian is chosen to be $H(1,1)=H_\mathrm{T}$.  At the initial stage of annealing $(s=0)$, since $\sigma_x\ket{+}_i=\ket{+}_i$,
the ground state $\ket{\psi_\mathrm{QP}}$ is a superposition of all possible $2^N$ states with an equal probability weight

\begin{equation} \label{QPp}
    \ket{\psi_\mathrm{QP}} :=\prod_i^N\ket{+}_i 
    =\frac{1}{\sqrt{2^N}}(\ket{\uparrow\uparrow\cdots\uparrow}+\ket{\uparrow\uparrow\cdots\uparrow\downarrow}+\cdots+\ket{\downarrow\downarrow\cdots\downarrow})\,,
\end{equation}
which we call the quantum paramagnetic (QP) phase. Here $\ket{\uparrow}$ and $\ket{\downarrow}$ denote eigenstates of $\sigma_z$ at each site, i.e., $\sigma_z \ket{\uparrow}_i = \ket{\uparrow}_i$ and $\sigma_z \ket{\downarrow}_i = -\ket{\downarrow}_i$.
\if{
To evaluate the required computational time (for this purpose, it is convenient to redefine $s=t/t_*, t\in[0,t_*]$), we refer to the adiabatic theorem. According to the adiabatic theorem, the computational time $t_*$ that is needed to efficiently obtain the ground state is proportional to the inverse square of the minimal energy gap $\Delta$ between the ground state and the first excited state ($t_*\sim\frac{1}{\Delta^2}$). For a large $N$, $\Delta$ is proportional to either $N^{-a}~(a>0)$ or $e^{-bN}~(b>0)$. (In the limit of $N\to \infty$, $\Delta$ goes to 0.) And by a lot of examples, it is known that $\Delta$ decays polynomially if a phase transition is second-order, whereas $\Delta$ decays exponentially if it is first-order \cite{Damski_2013,PhysRevB.71.224420,ikeda_universal_2020}. Therefore, the problem on system with second-order phase transition is efficiently solved. Moreover it is also known that when a non-stoquastic Hamiltonian is used, the first-order phase transitions can be avoided. The following $p$-spin model is such an example. 
}\fi
 The ground state of the target Hamlitonian $H_\mathrm{T}$ is degenerated if $p$ is even so we consider the cases when $p$ is odd. Then the ground state of $H_\mathrm{T}$ consists of all spins pointing up
 \begin{equation} \label{Fp}
   \ket{\psi_\mathrm{F}} :=  \ket{\uparrow\uparrow\cdots\uparrow} \,,
 \end{equation} 
which is the ferromagnetic (F) phase.
 
 Hence, as $s$ increases, a phase transition from QP phase to F phase occurs.   The corresponding order parameters are magnetizations 
 \begin{equation} \label{eq-mx}
     m_x := \frac{1}{N}\sum_{i}^N \left\langle X_i \right\rangle \,, \qquad m_z :=\frac{1}{N}\sum_{i}^N \langle Z_i\rangle  \,,
 \end{equation}
where the expectation value is obtained by the ground state at a given $s$. The first-order (second-order) phase transitions are defined by the discontinuity (continuity) of a given order parameter, respectively.  Without the antiferromagnetic interactions ($\lambda=1$), it is known that this model costs exponentially long time to obtain the ground state of $H_\mathrm{T}$ due to a first-order phase transition \cite{J_rg_2010}. The first-order phase transition can be avoided by strong enough antiferromagnetic interactions, with which the problem can be solved efficiently \cite{2012PhRvE..85e1112S,PhysRevA.95.042321,ikeda_universal_2020}.    In fact, by studying the behavior of $m_x$ along the quantum annealing process, the authors of \cite{2012PhRvE..85e1112S} showed that, for $p>3$, the system displays a first-order phase transition when $\lambda \gtrsim 0.4$, but the transition becomes second order for smaller values of $\lambda$. In their analysis, they consider a large $N$ approximation. This phase transition was also studied at large $N$ using a spin-coherent state technique \cite{PhysRevA.95.042321}, which gives results that are consistent with the analysis of \cite{2012PhRvE..85e1112S}.

In this section, we first reproduce some of the results of \cite{2012PhRvE..85e1112S} at {\it finite} $N$ without using a large $N$ approximation. By performing a direct diagonalization of the Hamiltonian (\ref{eq-hamQA}), we numerically compute the ground state expectation values in (\ref{eq-mx}). This is an important step because we will compute OTOCs at finite $N$ in the subsequent sections and try to understand its large $N$ properties. Thus, we first need to find an optimal $N$ which can be taken as large enough with reasonable computing time. 

\begin{figure}[]
\centering
\begin{subfigure}[]{0.45\hsize}
\includegraphics[width=\hsize]{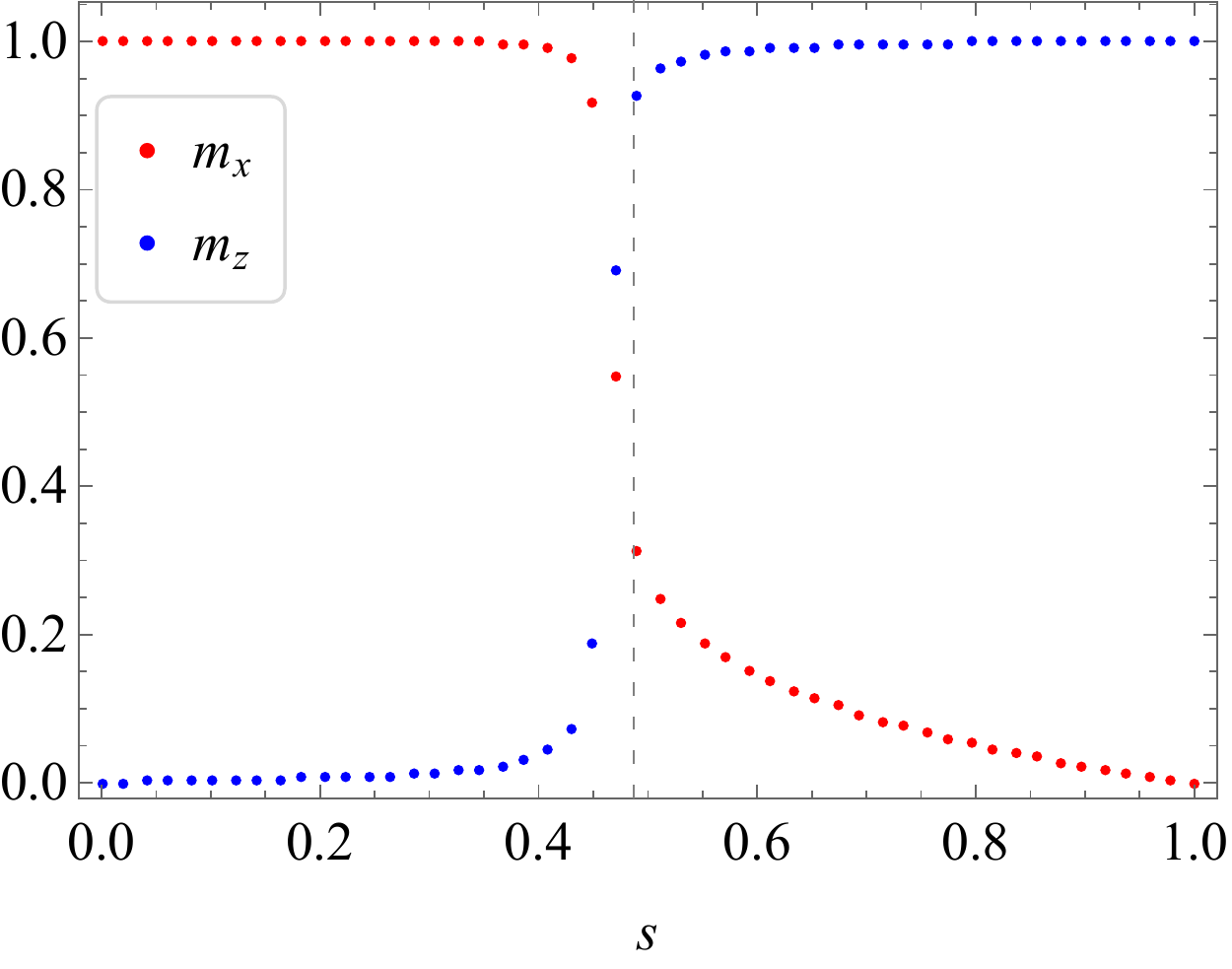}
    \caption{$N=11$, $p = 11$, $\lambda = 1.0$}
    \label{fig1:suba}
\end{subfigure} \ \ \
\begin{subfigure}[]{0.45\hsize}
\includegraphics[width=\hsize]{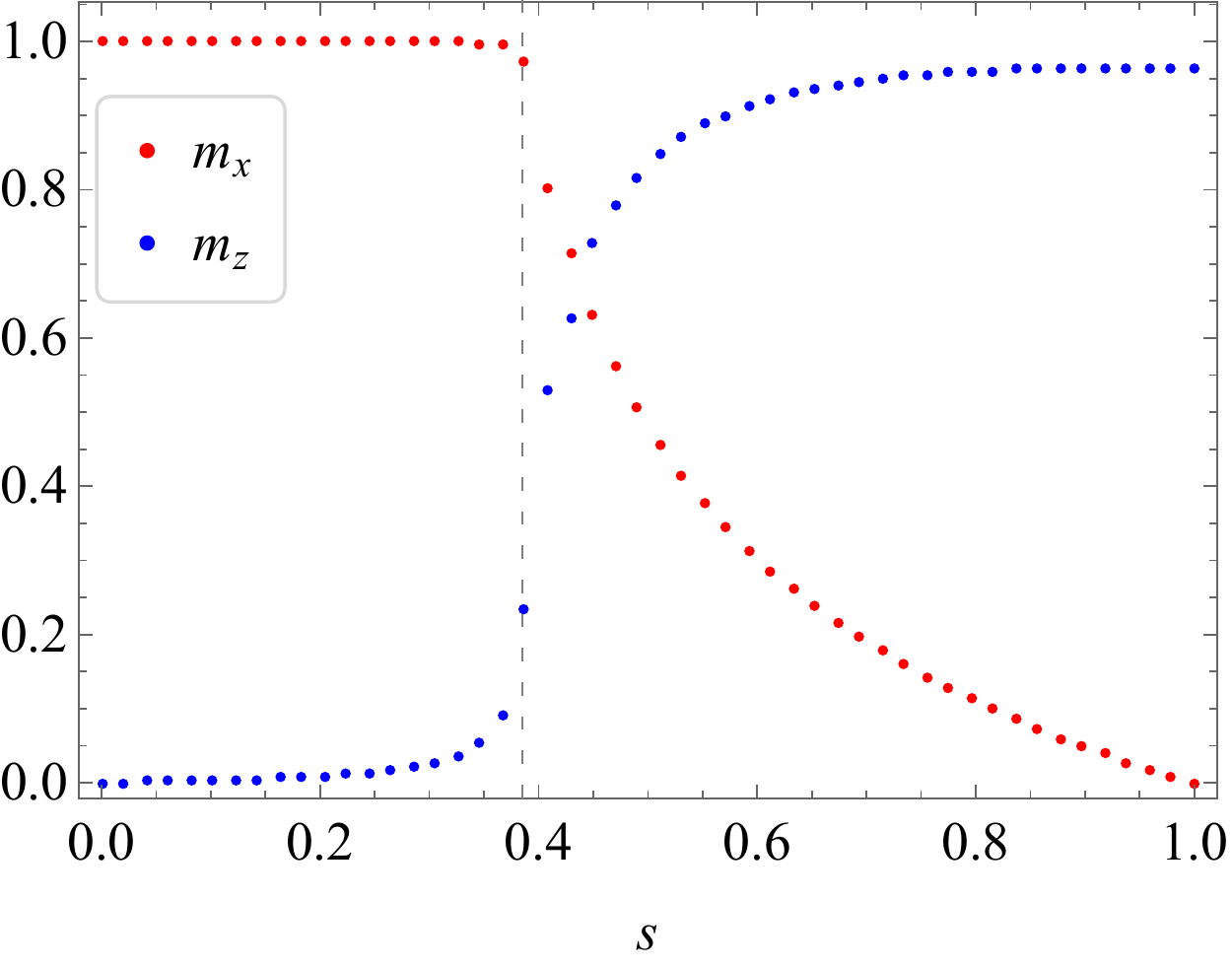}
    \caption{$N=11$, $p = 5$, $\lambda = 0.2$}
\end{subfigure}
    \caption{Magnetizations $m_x$ and $m_z$ versus $s$ for $N=11$. There are the first order(a) or second order(b) phase transitions between the QP phase \eqref{QPp} and the F phase \eqref{Fp}. The vertical dashed line indicates the transition point.}
    \label{fig:mag}
\end{figure}
\begin{figure}[]
\centering
\begin{subfigure}[]{0.45\hsize}
\includegraphics[width=\hsize]{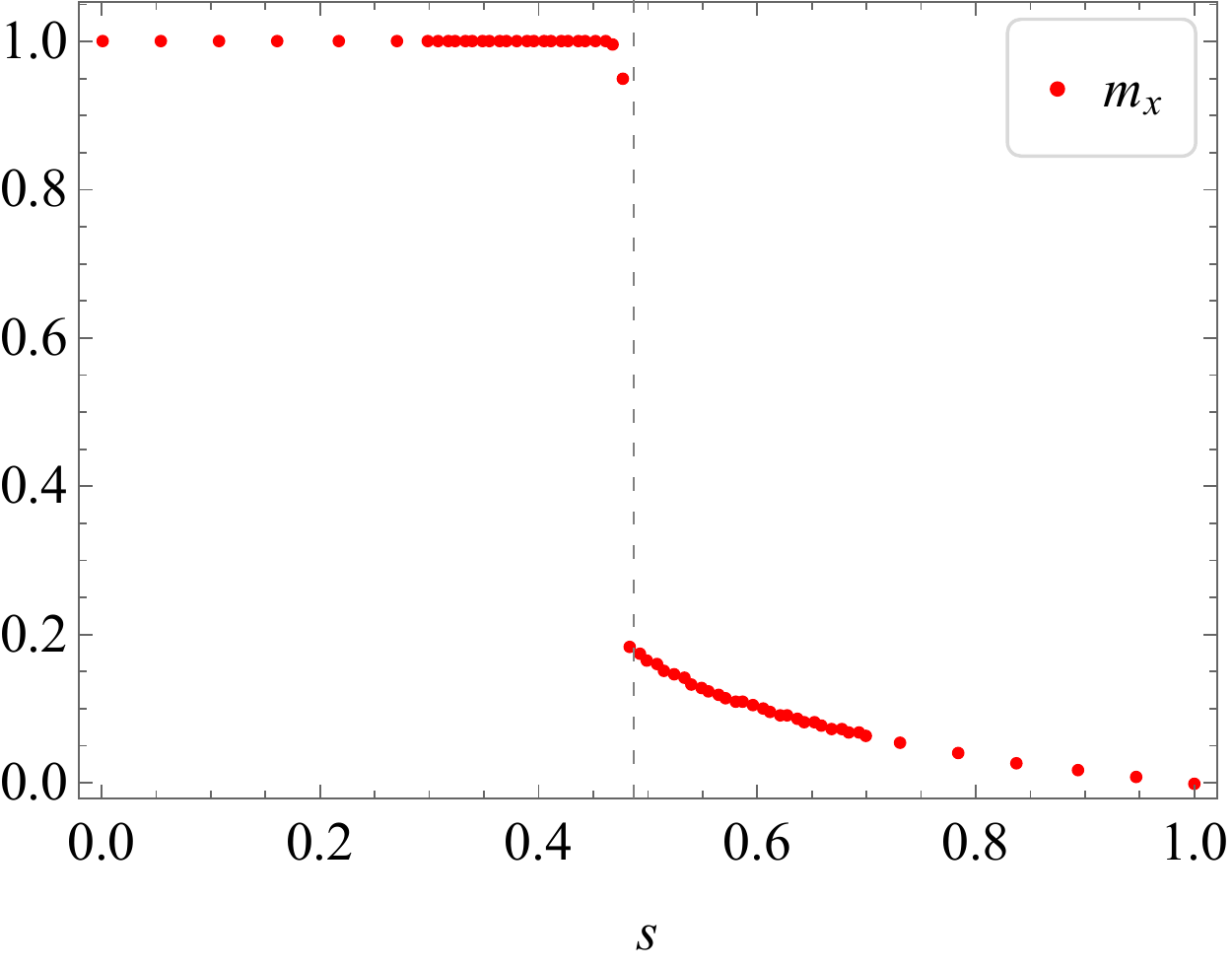}
    \caption{$N=20$, $p = 11$, $\lambda = 1.0$}
    \label{fig2:suba}
\end{subfigure} \ \ \
\begin{subfigure}[]{0.45\hsize}
\includegraphics[width=\hsize]{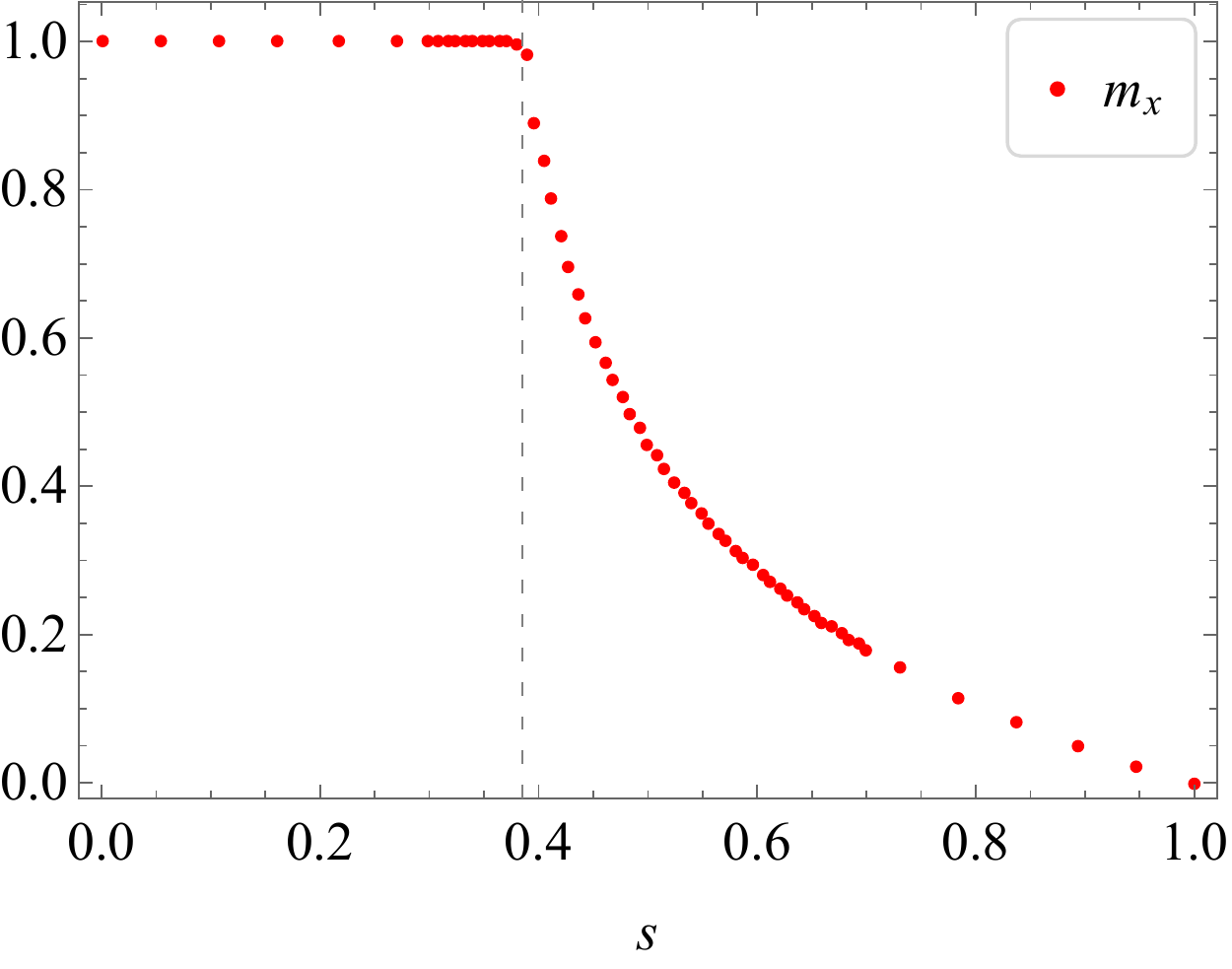}
    \caption{$N=20$, $p = 5$, $\lambda = 0.2$}
    \label{fig2:subb}
\end{subfigure}
    \caption{Magnetization $m_x$ versus $s$ for $N=20$. There are the first order(a) or second order(b) transitions between the QP phase \eqref{QPp} and the F phase \eqref{Fp}. The first order phase transition is clearer when compared to Fig.~\ref{fig1:suba}, where we set $N=11$. The vertical dashed line indicates the transition point. }
    \label{fig:mag1}
\end{figure}

Fig.~\ref{fig:mag} shows the behavior of the magnetizations $m_x$ and $m_z$ as functions of $s$ for some values of $N$, $\lambda$ and $p$. As expected, at $s=0$  $m_x$ ($m_z$) starts from one (zero) and decreases (increases) to zero (a constant positive value) as we increase $s$. We chose $N=11$ for the sake of comparison because it is the optimal value we will use in the following OTOC computation. If $\lambda \gtrsim 0.4$ the transition is first-order while if $\lambda \lesssim 0.4$ the transition is second-order. To show this we choose $\lambda =0.2$ and $\lambda = 1$. The feature of the first-order phase transition can be seen with $N=11$ in Fig.~\ref{fig1:suba}, but it is clearer for larger values of $N$. See Fig.~\ref{fig:mag1} where we set $N=20$.

\section{Diagnosing Quantum Phase Transition with OTOC}\label{sec:PT}
In the previous section we reviewed that the system described by the Hamiltonian (\ref{eq-hamQA}) displays a quantum phase transition from a quantum paramagnetic (QP) phase to a ferromagnetic (F) phase as we change the parameter $s$ from 0 to 1. In this section, we study quantum phase transitions from the viewpoint of quantum chaos, in particular in terms of out-of-time-order correlators (OTOCs). 

\subsection{Short Review on OTOC}

Here we provide a short review on OTOCs. For more details we refer to \cite{Jahnke:2018off}.\footnote{For recent applications in the context of holographic duality, see~\cite{Ahn:2019rnq,Jahnke:2019gxr} and references there in.} 
The OTOCs have the form
\begin{equation} \label{eq-OTOC}
    F_{V W}(t)=\langle W(t) V W(t) V \rangle\,, 
\end{equation}
where $V$ and $W$ are Hermitian and unitary local operators. In chaotic many-body systems, one expects $F_{V W}(t)$ to vanish at late times for almost any choice of operators $V$ and $W$. 

The late-time vanishing of OTOCs is tied  to the idea of scrambling of quantum information, which takes place in chaotic systems. Using the Baker-Campbell-Hausdorff (BCH) formula, we can write the Heisenberg operator as
\begin{equation}
    W(t)=e^{i H t} W e^{-i H t} = W+it\, [H,W]+\frac{(it)^2}{2!}[H,[H,W]]+ \cdots\,. 
\end{equation}
At $t=0$, the operator only involves the local degrees of freedom associated with $W$. Under time evolution, higher order terms in the BCH become important, and $W(t)$ becomes more and more complicated as it starts to act non-trivially in a increasing number of degrees of freedom. In other words, the operator $W(t)$ `grows' with time, and the initially local information gets scrambled into a non-local form.

The scrambling of the operator $W(t)$ can be probed by considering its commutator with some other, local operator, $V$. To avoid phase cancellations, one usually defines the double commutator
\begin{equation}
    C(t)= \langle |[W(t),V]|^2 \rangle\,,
\end{equation}
which starts at zero and grows as $W(t)$ scrambles with an increasing number of degrees of freedom. After the so-called `scrambling time', the operator $W(t)$ is scrambled with essentially all the degrees of freedom of the system, and $C(t)$ saturates to a constant value, which equals $2\, \langle V V \rangle \langle W W \rangle$. The double commutator is closely related to OTOCs. For unitary $V$ and $W$, the double commutator can be written as $C(t) = 2(1-\text{Re}[F_{VW}(t)])$. Hence, the saturation of the double commutator after the scrambling time implies the vanishing of OTOCs.\footnote{For an intuitive explanation (in the context of spin chains) of why the vanishing of OTOCs implies chaos, we refer to \cite{Huang:2017fng}. See also section 3 of \cite{Jahnke:2018off}.} In contrast, in non-chaotic systems, one expects a non-universal behavior of the OTOCs, which in general depends on the choice of the operators $V$ and $W$. Moreover, the absence of thermalization in integrable systems is expected to lead to an oscillatory behavior of OTOCs.

The way $F_{VW}(t)$ approaches to zero is particularly simple in systems that display some sort of classical limit (e.g.: quantum mechanical systems that have a well-defined $\hbar \rightarrow 0 $ limit or certain large-$N$ systems). In those cases, in an appropriate time window, the double commutator grows exponentially, i.e., $C(t)\sim e^{\lambda_L t}$, which  is reminiscent of the divergence in the distance between initially nearby trajectories in the phase space of classically chaotic systems. In those cases, one can define a quantum Lyapunov exponent $\lambda_L$ characterizing the onset of chaos in the system. By contrast, in standard spin chains models with local interactions, $F_{VW}(t)$ does not have any sort of exponential behavior with time \cite{Khemani:2018sdn,Craps:2019rbj}, even for systems that are known to be strongly chaotic by other more conventional criteria for quantum chaos. This prevents the definition of a quantum Lyapunov exponent for those systems\footnote{The time scale at which the OTOC vanishes is called the scrambling time $t_*$. If $C(t) \sim e^{\lambda_L t}$ at early times, then $t_* \sim \log N_\text{dof}$, where $N_\text{dof}$ is the number of degrees of freedom per site. For spin chains with local interactions, one typically has $t_* \sim \mathcal{O}(1)$, which leaves no `room' for an exponential growth \cite{Khemani:2018sdn,Craps:2019rbj}.} (see, however, \cite{Gharibyan:2018fax}). Despite the absence of exponential behavior, one usually expects to distinguish chaotic and integrable systems by the late-time behavior of OTOCs, which vanishes in the case of many-body chaotic systems. 

In finite-size systems, OTOCs do not decay exactly to zero at late times. In fact, it has been shown that the residual late-time value of OTOCs provides useful insights into the chaotic dynamics \cite{Huang:2017fng}. For energy-conserving chaotic spin chains, for example, the late-time value of the OTOCs scale as an inverse polynomial in the system size \cite{Huang:2017fng}. 

\subsection{Ground state OTOCs} \label{sec-resultOTOC}

We are interested in ground state OTOCs
\begin{equation} \label{defgs}
    F_{VW}^{(0)}(t)=\bra{\phi_0}W(t) V W(t) V \ket{\phi_0}\,, 
\end{equation}
where $V$ and $W$ are Hermitian operators and $\ket{\phi_0}$ is the ground state of the total Hamiltonian $H=H(s,\lambda)$. In the following, we consider cases where $V$ and $W$ are non-local operators, like 
\begin{equation}
  S^x=\frac{1}{2}\sum_{i=1}^{N}\sigma^{x}_i  \,, \quad \mathrm{or} \quad S^z=\frac{1}{2}\sum_{i=1}^{N}\sigma^{z}_i \,,
\end{equation}
and also cases where $V$ and $W$ are local unitary operators, like $X_i$ or $Z_i$, for some site $i$ as in \eqref{xz12}. 

\begin{figure}[]
\centering
    \includegraphics[width=0.55\hsize]{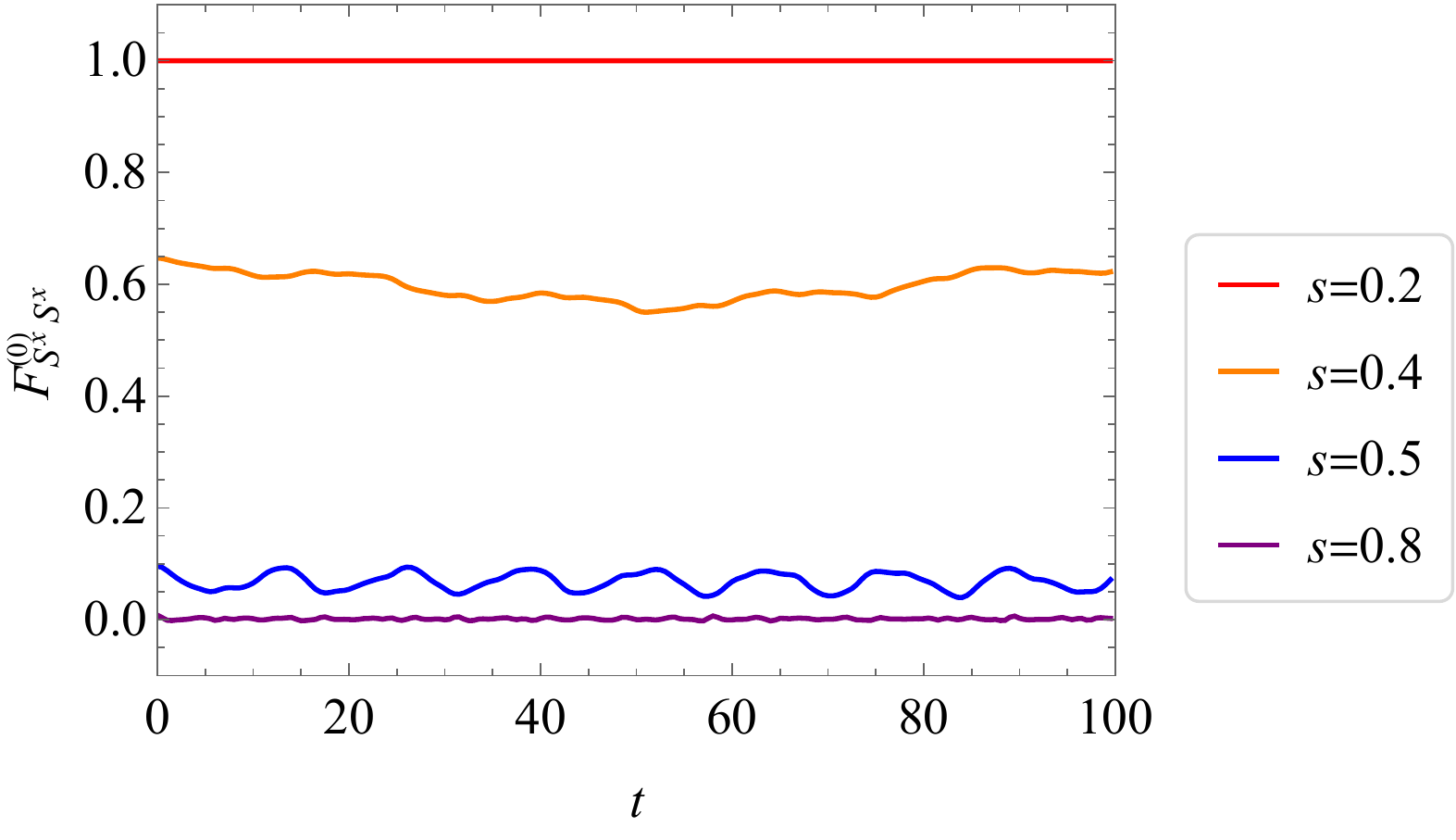}
\caption{Time dependence of $ F_{S^x S^x}^{(0)}(t)$ for increasing values of the annealing parameter $s$. Here we set $N=11$, $p=5$ and $\lambda=0.2$. The OTOC oscillates around a constant positive value that depends on the value of $s$. }
  \label{fig:OTOC}
\end{figure}

In both cases, the OTOCs show a qualitatively similar behavior, but the results for non-local operators tend to be more smooth. As an example, we show in Fig.~\ref{fig:OTOC} the time dependence of $F_{S^x S^x}^{(0)}(t)$ for increasing values of the annealing parameter $s$. 
%
In the beginning of the annealing process, when $s < s_c$ (in this case, $s_c = 0.385$) and the system is in the QP phase, $F_{S^x S^x}^{(0)}$ oscillates around one. After the phase transition to the F phase, for $s > s_c$,  $F_{S^x S^x}^{(0)}$ oscillates around a constant (smaller than one) value that quickly decreases to zero as $s$ approaches one.

\begin{figure}[]
\centering
\begin{subfigure}[]{0.45\hsize}
    \includegraphics[width=\hsize]{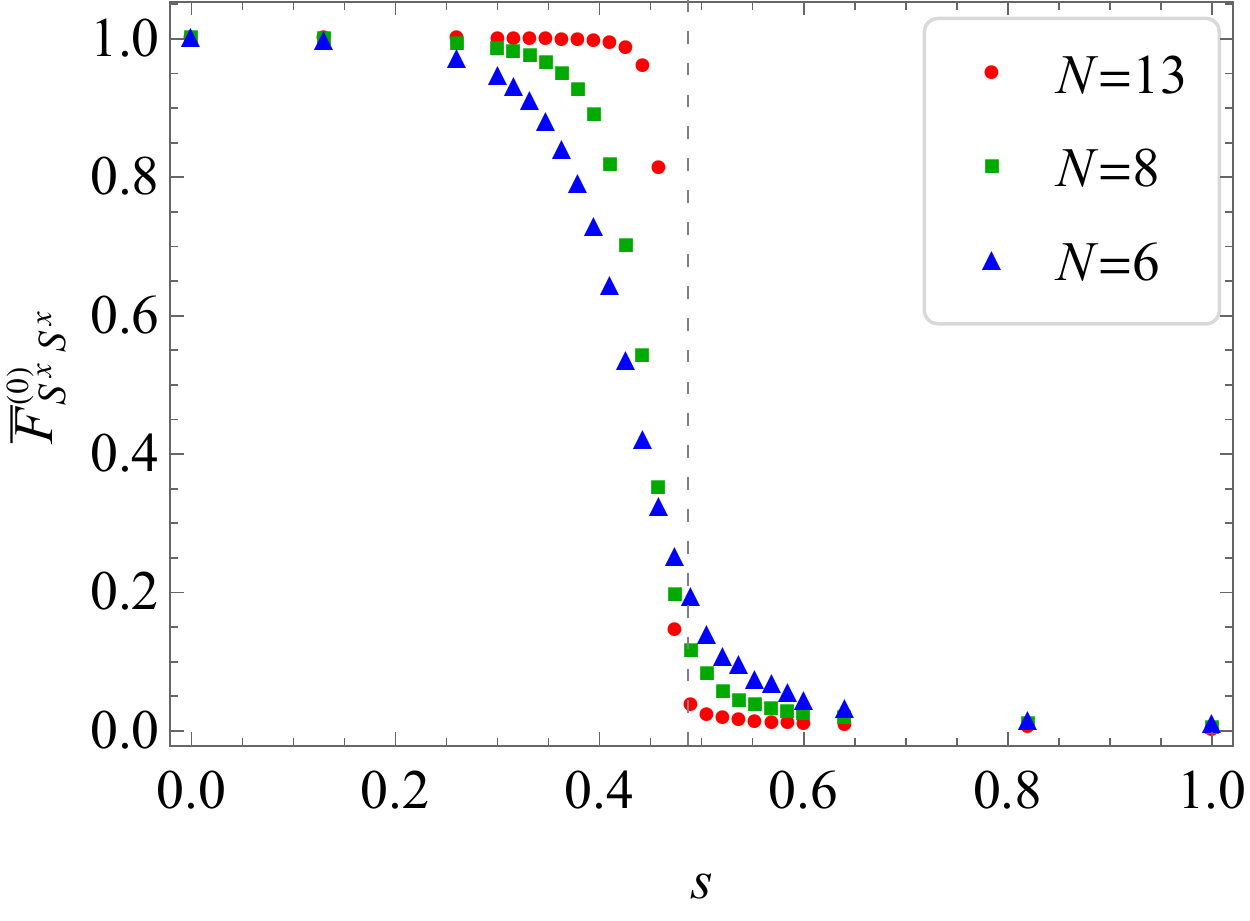}
    \caption{$p=11, \lambda=1.0$ }
    \label{fig4:suba}
\end{subfigure} \ \ \
\begin{subfigure}[]{0.45\hsize}
    \includegraphics[width=\hsize]{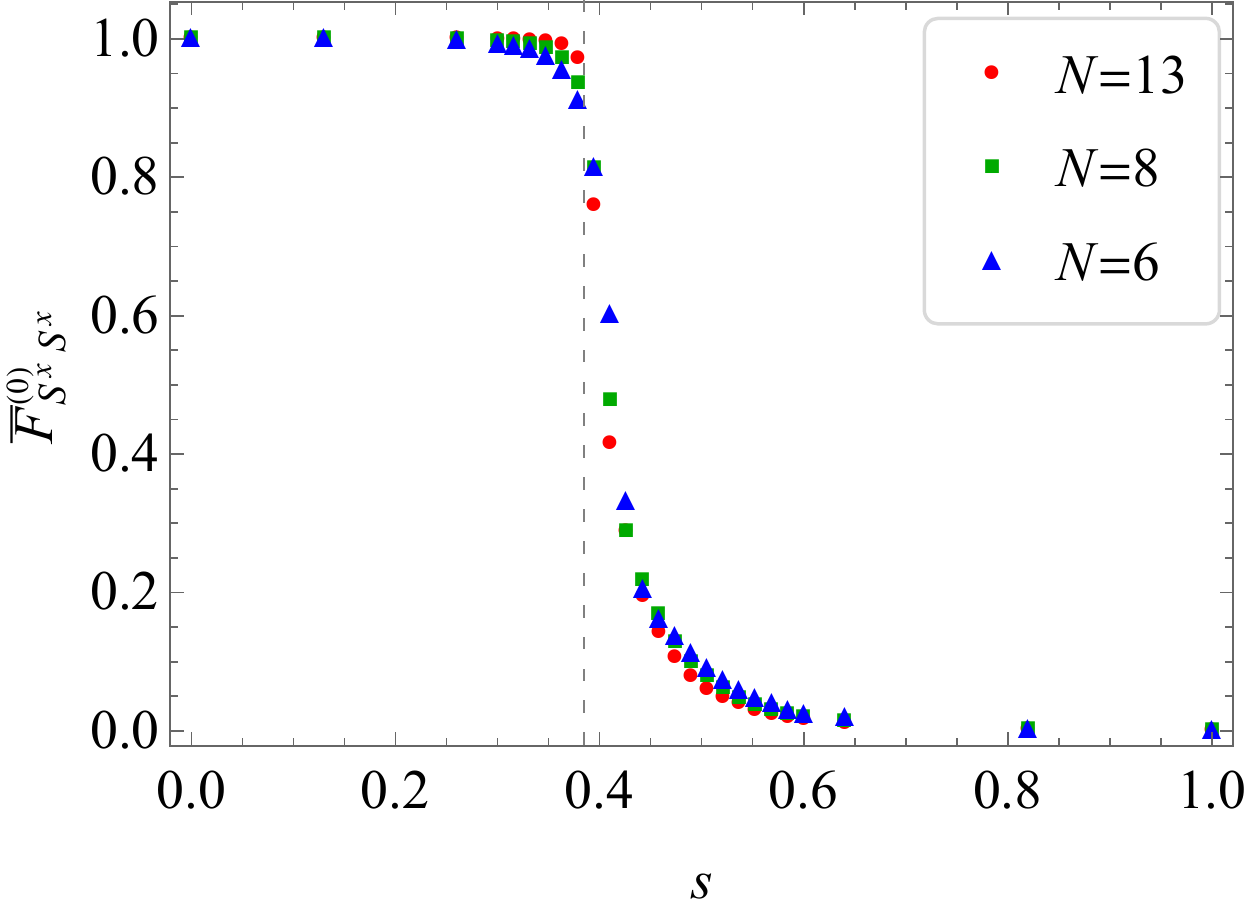}
    \caption{$p=5, \lambda=0.2$}
    \label{fig4:subb}
\end{subfigure}
\caption{Time-averaging OTOCs $\bar{F}_{S^x S^x}$ versus the annealing parameter $s$. Here we set $N=6,8,13$. The phase transition is first-order for $(p,\lambda)=(11,1.0)$ and second-order for $(p,\lambda)=(5,0.2$). The transition points of the second-order phase transitions obey the formula $s=\frac{1}{3-2\lambda}$ and those of the first-order phase transitions are numerically estimated in \cite{PhysRevA.95.042321}.}
    \label{fig:Fbar}
\end{figure}

The above time-behavior indicates that the time-averaging OTOCs
\begin{equation}
    \bar{F}_{V W}^{(0)}=\lim_{T\to\infty}\frac{1}{T}\int_0^T F_{V W}^{(0)}(t)\, dt\,,
\end{equation}
can diagnose the phase transition.
We have checked that this is indeed true by studying the behavior of $\bar{F}_{V W}^{(0)}$ along the annealing process. 
Fig.~\ref{fig:Fbar} shows the behavior of $\bar{F}_{S^x S^x}^{(0)}$ as a function of the annealing parameter $s$.
The time-averaging OTOC changes behavior across the quantum phase transition, whose transition points (estimated at large $N$) are indicated in the plots by vertical lines. The transition points of the second-order phase transitions obey the formula $s_c=\frac{1}{3-2\lambda}$ and those of the first-order phase transitions are numerically estimated in \cite{PhysRevA.95.042321}. Those estimates become more and more precise as we consider larger values of $N$.

The choice of parameters for Fig.~\ref{fig:Fbar} is such that we have a first-order phase transition in Fig.~\ref{fig4:suba}, where $(p,\lambda)=(11,1)$, and a second-order phase transition in Fig.~\ref{fig4:subb}, where $(p,\lambda)=(5,0.2)$.  The order of the phase transition is visible in the behavior of the time-averaging OTOC, specially when we consider the curves for $N=13$, where we can see that $\bar{F}_{S^x S^x}^{(0)}$ has a discontinuous (continuous) behavior at the quantum transition point in first-order (second-order) phase transitions. 

\begin{figure}[]
\centering
\begin{subfigure}[]{0.45\hsize}
    \centering
    \includegraphics[width=\hsize]{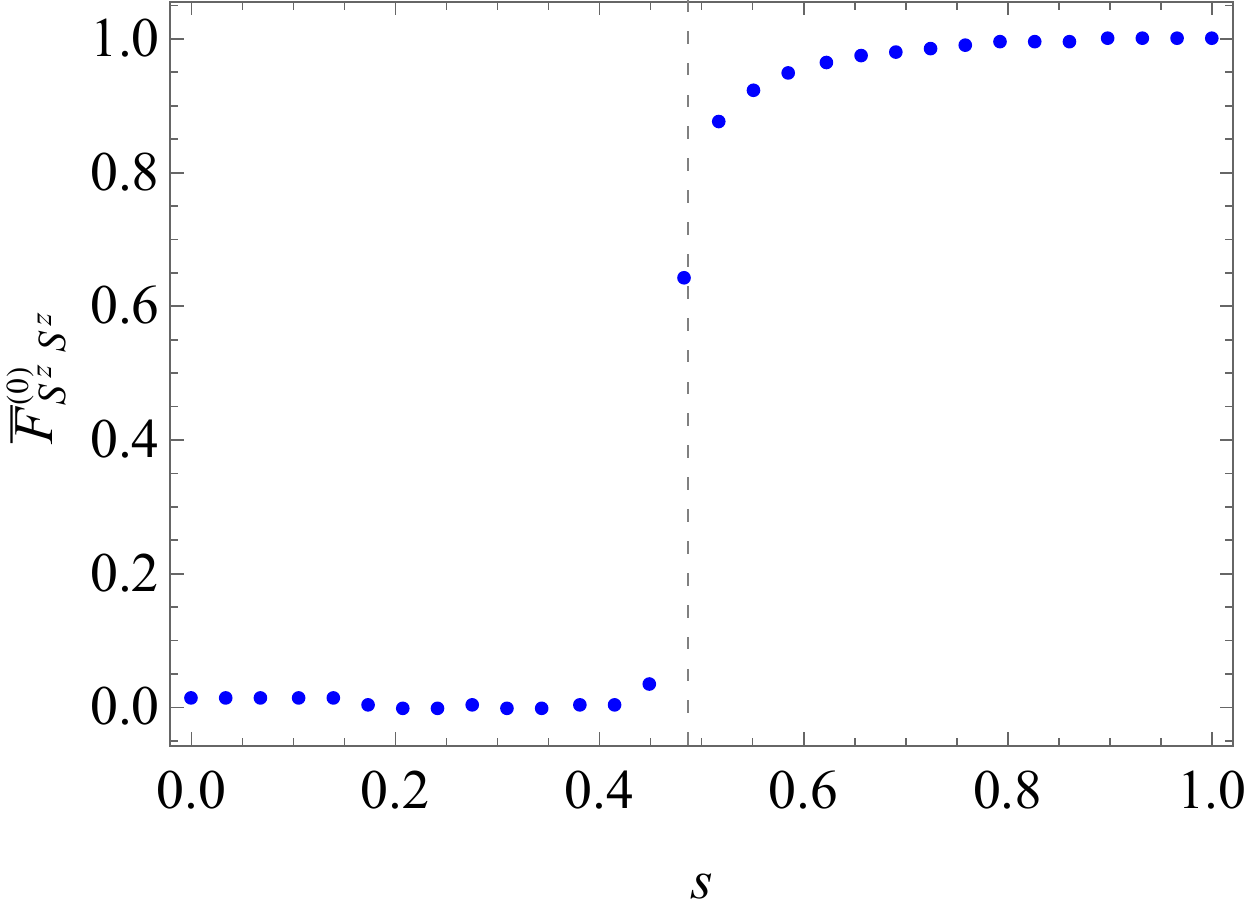}
    \caption{$p=11, \lambda=1.0$}
    \label{fig5:suba}
\end{subfigure} \ \ \
\begin{subfigure}[]{0.45\hsize}
    \centering
    \includegraphics[width=\hsize]{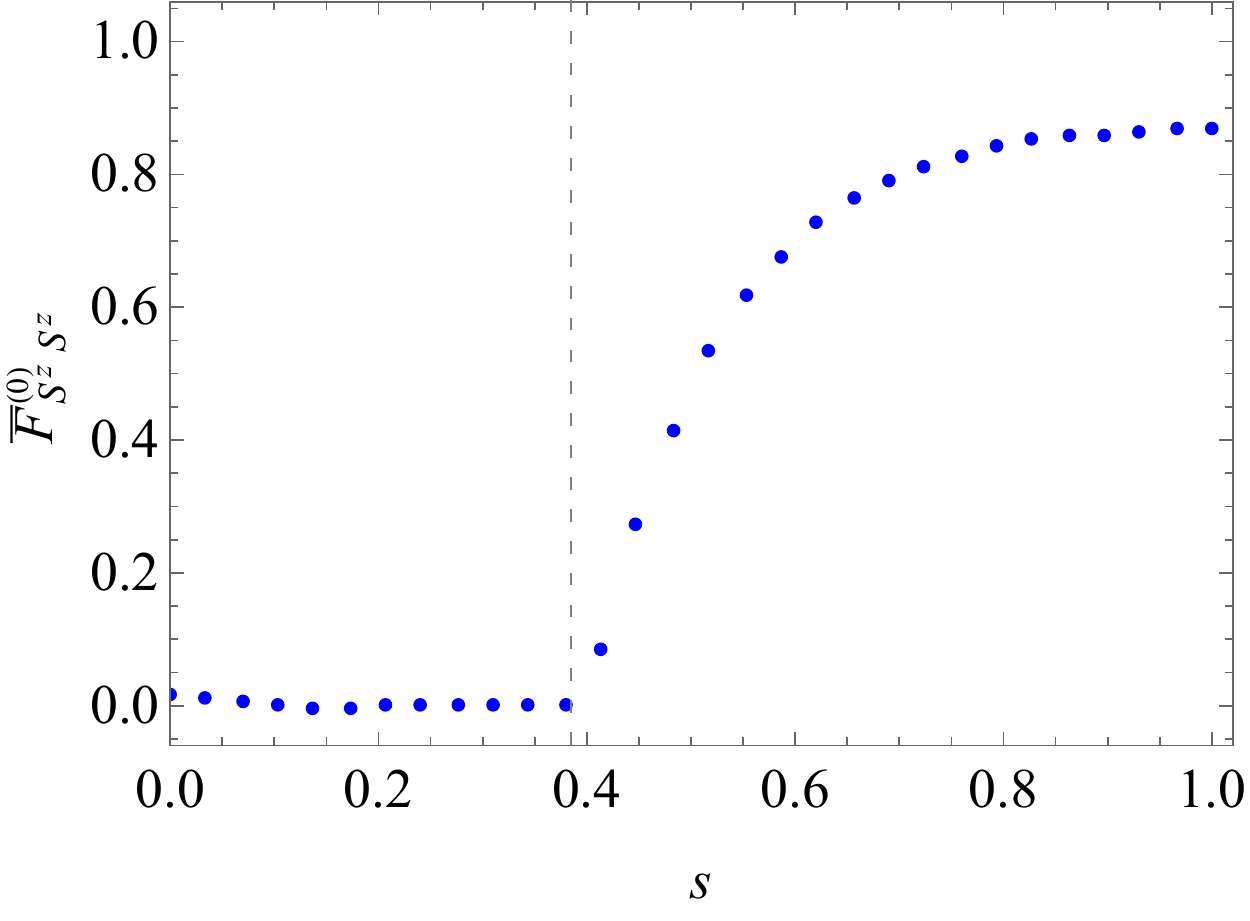}
    \caption{$p=5, \lambda=0.2$}
    \label{fig5:subb}
\end{subfigure}
    \caption{Time-averaging OTOC $\bar{F}_{S^z S^z}^{(0)}$ versus the annealing parameter $s$. The transition points for each case are indicated by vertical lines. Here we set $N=11$.}
    \label{fig:X}
\end{figure}

The diagnosis of the phase transition in terms of time-averaging OTOCs is not limited to a particular choice of operators.  Fig.~\ref{fig:X} shows that $\bar{F}_{S^z S^z}^{(0)}$ changes behavior around the quantum transition point, and also identifies the order of the phase transition. The same is true for time-averaging OTOCs involving local operators, e.g., $\bar{F}^{(0)}_{X_1 X_8}$, as it is shown in Fig.~\ref{fig:OTOC_sep}. We have  also checked that it is possible to detect the phase transition with different combinations, like $(V,W)= (Y_i,X_i),\, (X_i,Z_i)$, etc. In this sense, we may say that the OTOCs provide a generalization of the order parameter, and thereby are useful to characterize a phase transition when the rigorous order parameter is unknown.

\begin{figure}[]
\centering
\begin{subfigure}[]{0.45\hsize}
    \includegraphics[width=\hsize]{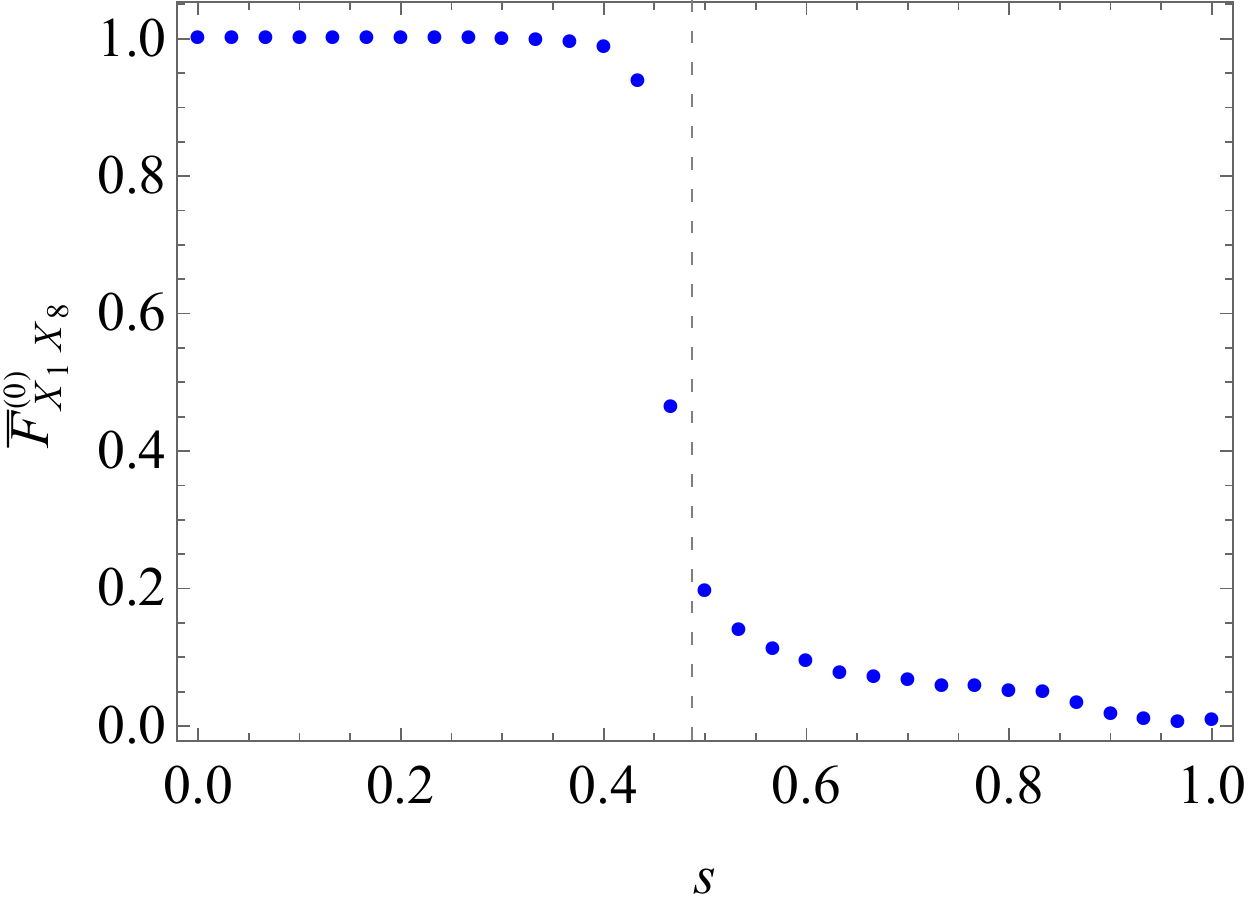}
    \caption{$p=11, \lambda=1.0$}
\end{subfigure} \ \ \
\begin{subfigure}[]{0.45\hsize}
    \includegraphics[width=\hsize]{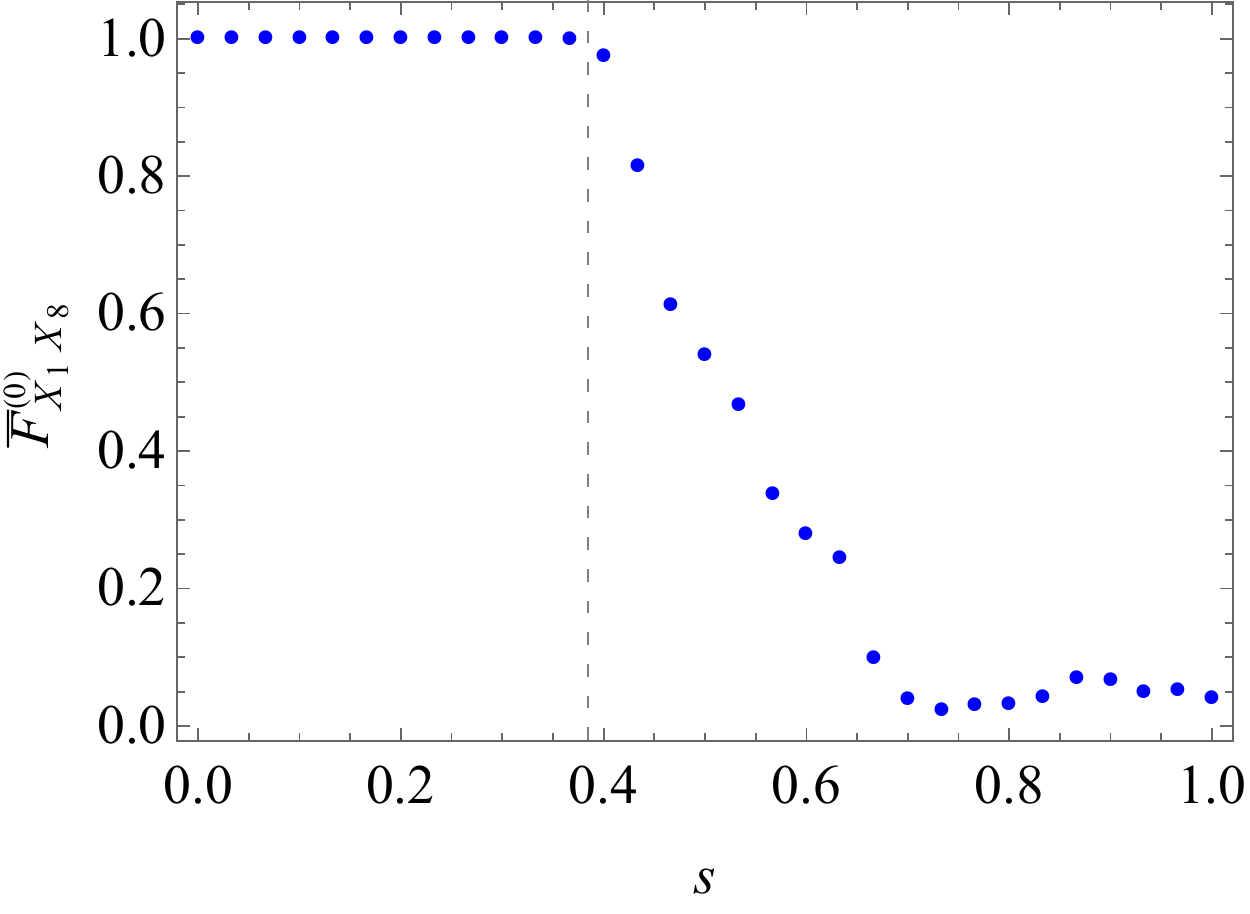}
    \caption{$p=5, \lambda=0.2$}
\end{subfigure}
    \caption{Time-averaging OTOC $\bar{F}_{X_1 X_8}^{(0)}$ versus the annealing parameter $s$. The transition points for each case are indicated by vertical lines. Here we set $N=11$.}
    \label{fig:OTOC_sep}
\end{figure}

Naively, the fact that $\bar{F}^{(0)}_{X_1 X_8}$ is initially one in the QP phase, and zero in the F phase, seem to suggest that the F phase is chaotic, but this is not correct. In fact, time-averaging OTOCs of the form $\bar{F}^{(0)}_{Z_i Z_j}$ start at zero in the QP phase, and approach one in the F phase. See Fig.~\ref{fig8:suba}. That means that the value that $\bar{F}^{(0)}_{V W}$ takes in each phase depends on the operators $V$ and $W$ in the corresponding OTOC. We understand this operator dependence of $\bar{F}^{(0)}_{V W}$ as due to the absence of scrambling in the ground state OTOCs. As we will see in the next section, the system under consideration displays some degree of scrambling at finite temperature, which makes thermal OTOCs to oscillate around zero at late-times in both phases. However, as we reduce the temperature, the physics is dominated by the ground state, and the behavior of $\bar{F}^{(0)}_{Z_i Z_j}$ and $\bar{F}^{(0)}_{X_i X_j}$ as a function of $s$ starts to mimic the behavior of the magnetizations $m_z$ and $m_x$, respectively.

\if{
\begin{figure}[H]
\begin{minipage}{0.49\hsize}
    \includegraphics[width=\hsize]{i1_j8_p11_L8_lambda1_ver2.pdf}
\end{minipage}
\begin{minipage}{0.49\hsize}
    \includegraphics[width=\hsize]{XZ_i1_j8_p11_L8_lambda1.pdf}
\end{minipage}
    \caption{Time-averaging OTOC $\bar{F}_{V W}^{(0)}$ versus the annealing parameter $s$. The case with $V=Z_N$ and $W=Z_1$  is shown on the left panel, while the case with $V=Z_N, W=X_1$ is shown on the right panel. We set $N=8$ and $\lambda=1$.}
    \label{fig:OTOC_sep}
\end{figure}
}\fi

\subsection{Thermal OTOCs} \label{sec-thermalOTOC}
We now consider thermal OTOCs to study thermal effects and the relation to chaos.  
Similarly to the ground state OTOCs, we first study the time-averaging thermal OTOC 
\begin{equation}
  \bar{F}_{S^x S^x}^{\beta} \equiv \lim_{T \rightarrow \infty} \frac{1}{T}\int_{0}^{T} F_{S^x S^x}^{\beta}(t) dt\,,
\end{equation}
where 
\begin{equation}
    F_{S^x S^x}^{\beta}(t) \equiv \langle F_{S^x S^x}(t) \rangle_{\beta} \equiv \frac{\text{Tr} \left[ e^{-\beta H} F_{S^x S^x}(t) \right]}{\text{Tr} \left[ e^{-\beta H} \right]} \,,
\end{equation}
with the inverse temperature $\beta$.

\begin{figure}[]
\centering
\begin{subfigure}[]{0.45\hsize}
    \includegraphics[width=\hsize]{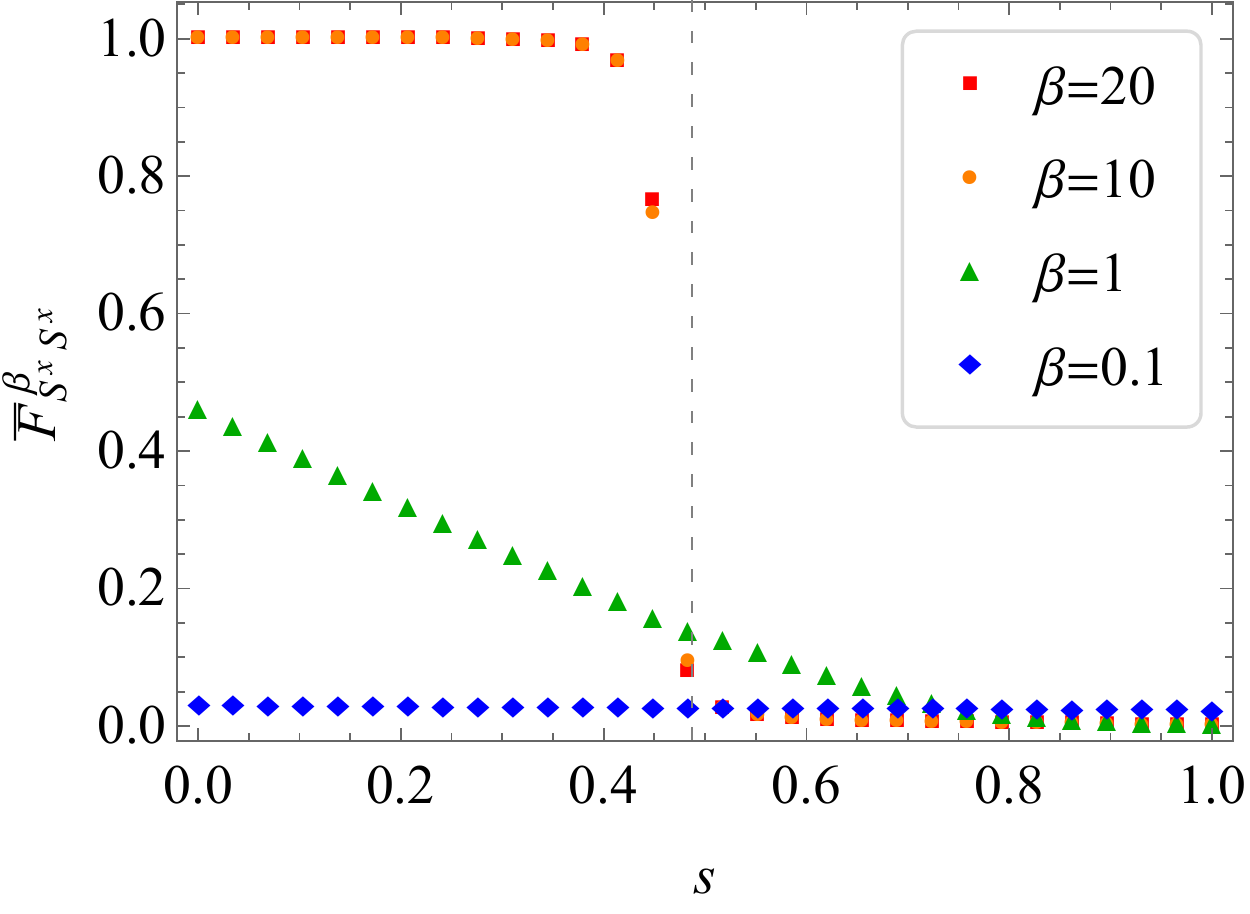} 
    \caption{$p=11, \lambda=1.0$}
\end{subfigure} \ \ \
\begin{subfigure}[]{0.45\hsize}
    \includegraphics[width=\hsize]{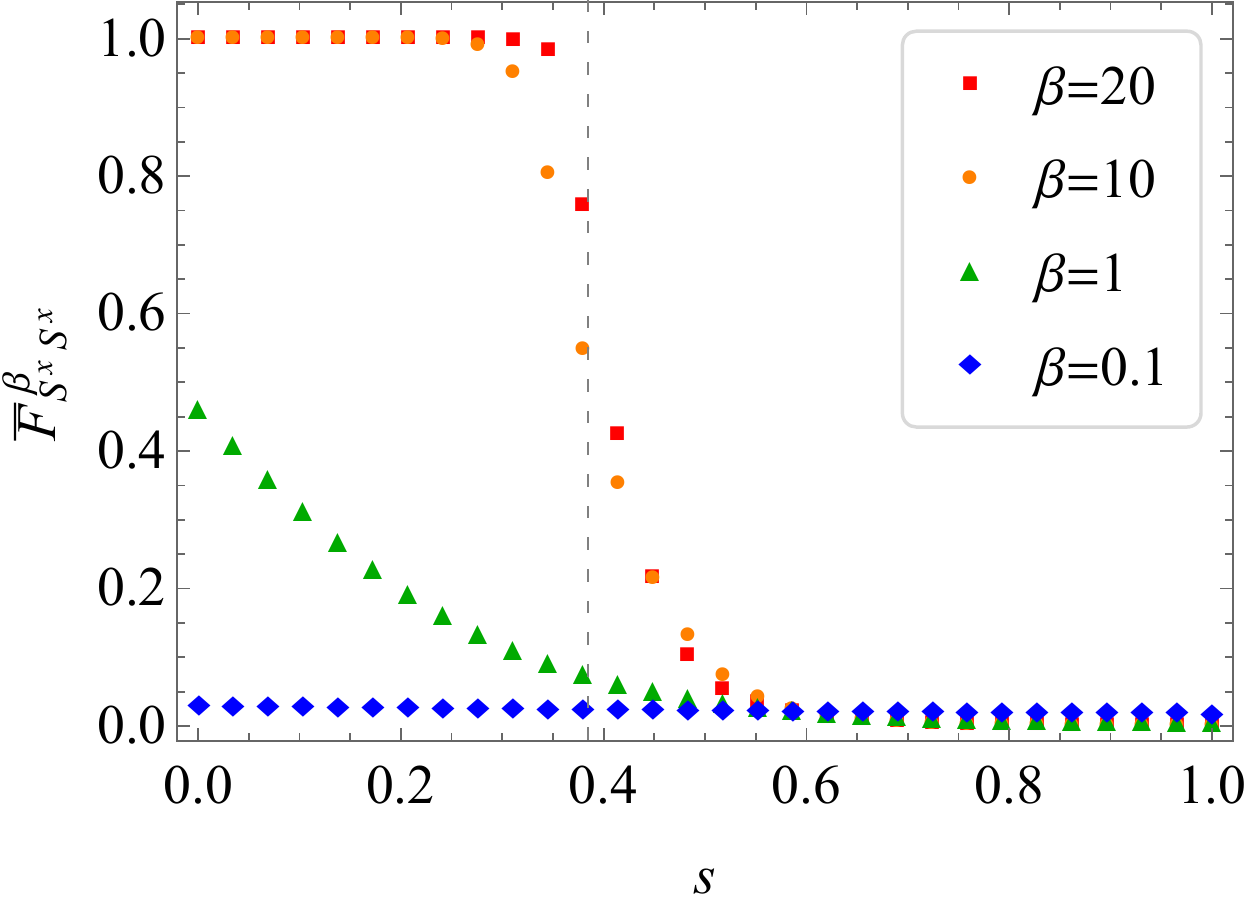}
    \caption{$p=5, \lambda=0.2$}
\end{subfigure}
    \caption{ Time average of the thermal OTOC, $\bar{F}_{S^x S^x}^{\beta}$ versus the annealing parameter $s$.  The transition points for each case are indicated by vertical lines. We set $N$ = 11. $\beta$ is the inverse temperature.}
    \label{fig:TOTOC}
\end{figure}

Fig.~\ref{fig:TOTOC} shows that the time-averaging thermal OTOC $ \bar{F}_{S^x S^x}^{\beta}$ can successfully diagnose the phase transition at small temperatures as expected. In high temperature regimes, $ \bar{F}_{S^x S^x}^{\beta}$ vanishes in both phases.
At small temperatures, thermal OTOCs are dominated by a contribution coming from the ground state, and contributions from higher excited states are exponentially suppressed. In fact, if $\mathcal{F}(t)= W(t) V(0) W(t) V(0)$, we can write
\begin{equation}\label{eq-OTOCth}
    \langle \mathcal{F}(t) \rangle_{\beta}=\frac{\text{Tr} \left[ e^{-\beta H} \mathcal{F}(t) \right]}{\text{Tr} \left[ e^{-\beta H} \right]} =\frac{\sum_{n=0}^{\infty}e^{-\beta E_n} \langle n|\mathcal{F}(t)|n\rangle}{\sum_{n=0}^{\infty}e^{-\beta E_n}}= \langle 0|\mathcal{F}(t)|0\rangle+\sum_{n=1}^{\infty}\mathcal{O}\left( e^{-\beta(E_n-E_0)}\right)\,,
\end{equation}
where $|n \rangle$ denotes the state with energy $E_n$, with $E_0$ being the ground state energy. From (\ref{eq-OTOCth}) we see that corrections from excited states are all very small as long as the temperature is much smaller than the difference in energy between the ground state and the first excited state, i.e., $\beta (E_1-E_0)>> 1$. Therefore, at small temperatures, the thermal OTOC is roughly the same as the ground state OTOC, and that is why it can also diagnose phase transitions. For a more details about the relation between ground state OTOCs and phase transitions, we refer to \cite{Dag:2019yqu}.

\begin{figure}[]
\centering
\begin{subfigure}[]{0.45\hsize}
    \includegraphics[width=\hsize]{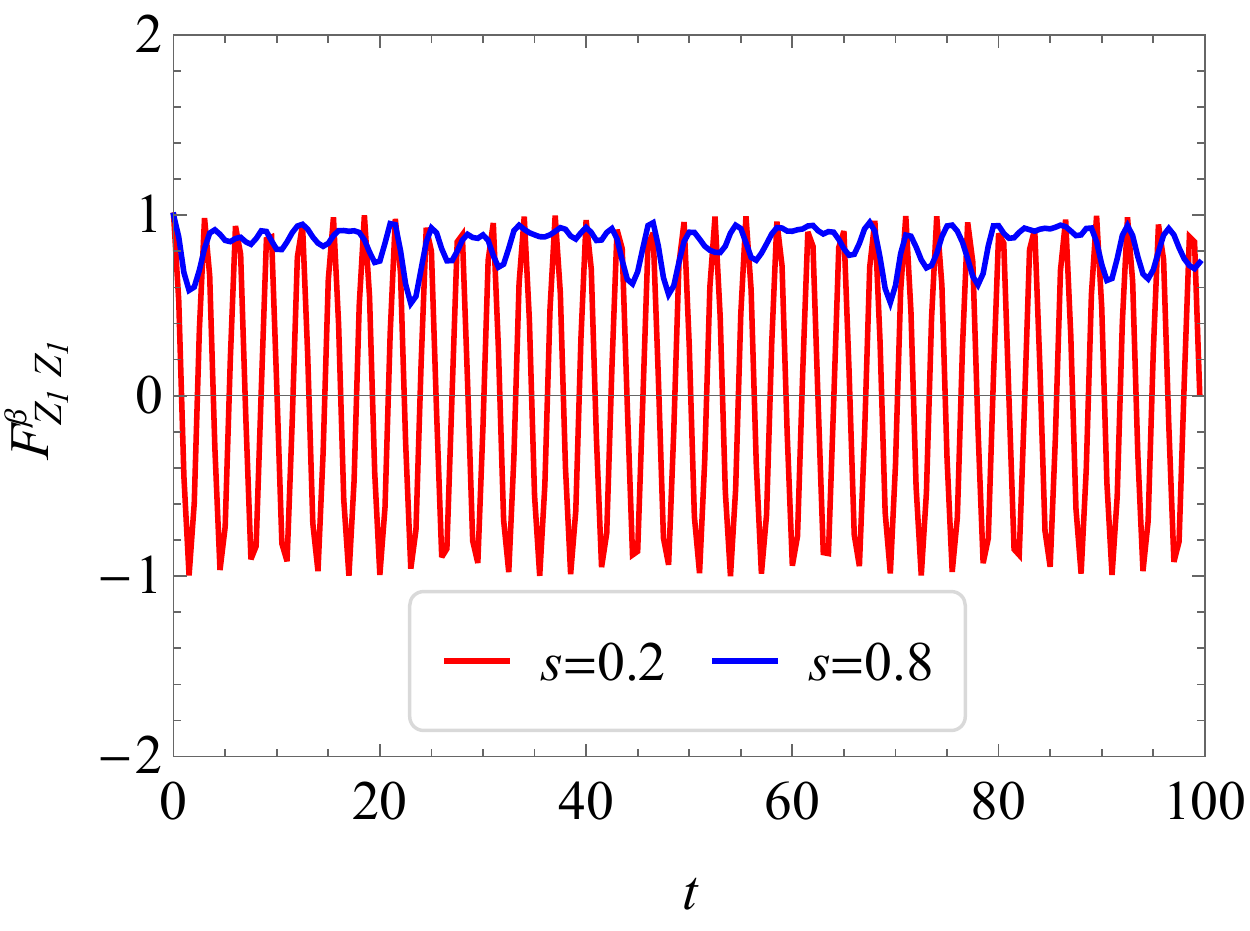}
    \caption{Small temperature ($\beta=100$)}
    \label{fig8:suba}
\end{subfigure}
\hspace{0.1cm}
\begin{subfigure}[]{0.45\hsize}
    \includegraphics[width=\hsize]{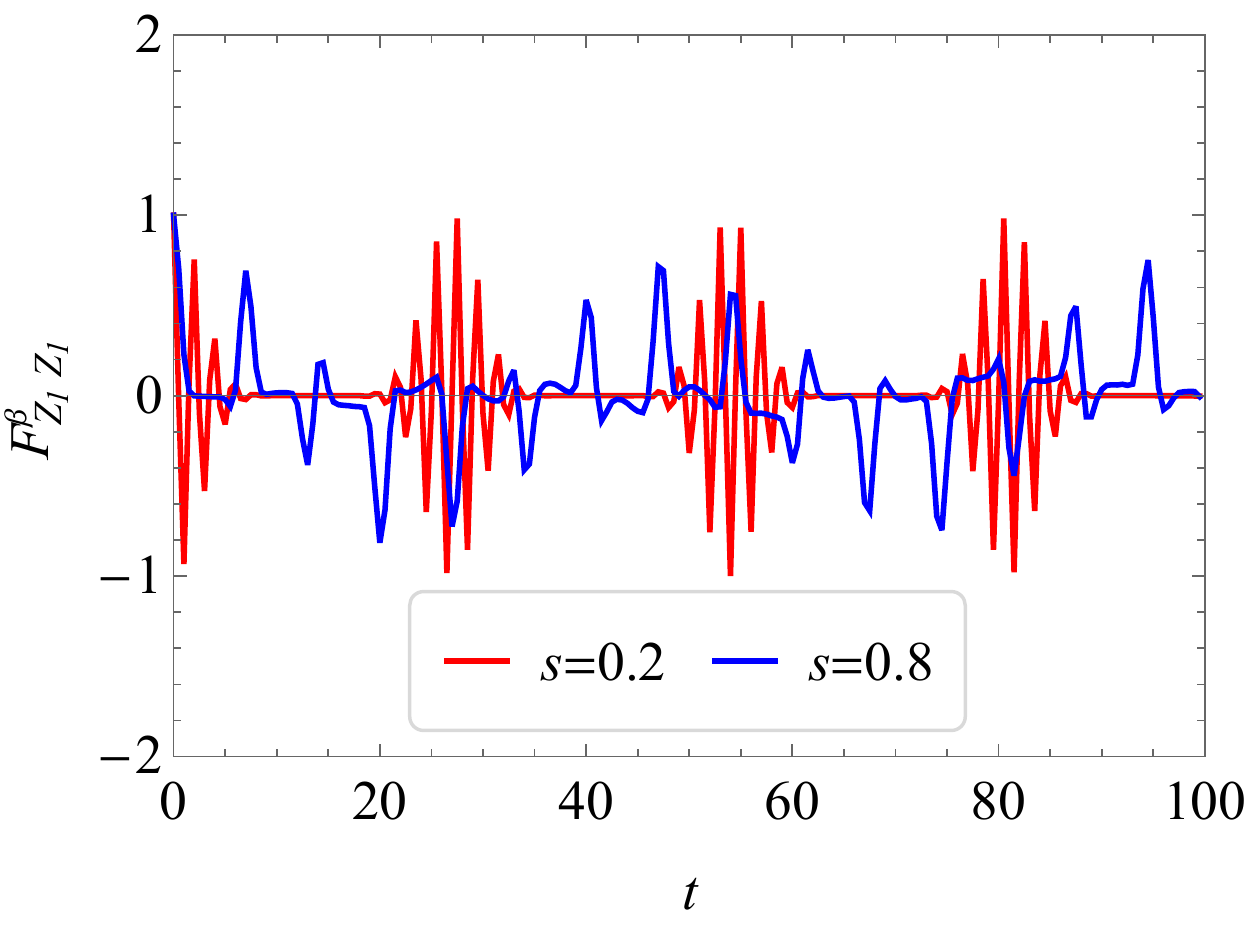}
    \caption{Infinite temperature ($\beta=0$)}
    \label{fig8:subb}
\end{subfigure}
\caption{Time dependence of $ F_{Z_i Z_j}^{\beta}$ for $N=11$, $p=5$ and $\lambda=0.2$. The red curves are for $s=0.2$ (QP phase) and the blue curves are for $s=0.8$ (F phase).}
  \label{fig:OTOCthermal}
\end{figure}

Next, to have a clear understanding of the scrambling properties of our model, we also consider thermal OTOCs of the form, without time averaging, 
\begin{equation}
F_{Z_i Z_j}^{\beta}(t) = \frac{\text{Tr}\,[ e^{-\beta H} Z_i(t) Z_j(0) Z_i(t) Z_j(0) ]}{\text{Tr}[e^{-\beta H}]}\,,
\end{equation}
where $\beta$ is the inverse temperature and $H$ is given by (\ref{eq-hamQA}). Fig.~\ref{fig:OTOCthermal} shows the time dependence of $F_{Z_1 Z_1}^{\beta}$
for the system in the QP phase ($s=0.2$, red curves), and for the system in the F phase ($s=0.8$, blue curves). In Fig.~\ref{fig8:suba}, we consider a very small temperature by setting $\beta=100$. In this case, the thermal OTOC oscillates around  a constant value which is close to one (zero) in the QP (F) phase. In Fig.~\ref{fig8:subb}, we show the results at infinite temperature ($\beta=0$). In this case, the thermal OTOC oscillates around zero in both phases.

In chaotic systems, thermal OTOCs are expected to approach zero at late times. In our case, thermal OTOCs computed at infinite temperature ($\beta=0$) display an oscillatory behavior that persists at very large times, oscillating around zero and taking values that roughly range between -1 and 1. See Fig.~\ref{fig8:subb}. The oscillatory behavior around zero is present in both phases of the annealing process, and it happens independently of the operator considered in the OTOC. This should be contrasted with the non-universal behavior of ground state OTOCs $F_{VW}^{(0)}(t)$, which oscillate around a constant value that depends both on the phase and on the operators $V$ and $W$.

We interpret the fact that all thermal OTOCs oscillate around zero at high temperatures as some sort of weak chaotic behavior of the model (\ref{eq-hamQA}). The adjective {\it weak} is because the thermal OTOC do not vanish at late-time, but only oscillates around zero. We believe this characterizes a weak chaotic behavior that takes place in relatively simple systems at finite temperature. We provide a more detailed discussion about the idea of weak scrambling in Appendix \ref{sec:app}.

\section{Level Spacing Statistics}\label{sec:LS}
 The statistics of the spacing between consecutive energy eigenvalues (level spacing) differs significantly depending on whether we consider chaotic or integrable systems. In chaotic systems, the level spacing statistics obeys a Wigner-Dyson distribution \cite{BGS}, while in integrable systems it follows an Poisson distribution \cite{BTZ-Poisson,Berry-80}. The above statement holds for a properly normalized spectrum, in which the influence of model-dependent density of states is removed by normalizing the level spacing by the local density of states. A properly normalized spectrum is said to be unfolded. Moreover, if the Hamiltonian has symmetries, one needs to block-diagonalize it according the corresponding to conserved charges, and perform the unfolding for each block, because the eigenvalues of different blocks are uncorrelated. 

The Hamiltonian (\ref{eq-hamQA}) conserves the total spin operator ${\bf S}^2=(S^x)^2+(S^y)^2+(S^z)^2$, where $S^x=\frac{1}{2}\sum_{i=1}^{N}\sigma^{x}_i$, and so on. We will be interested in the sector with maximal total spin, in which $S=N/2$, where $S$ is the angular momentum quantum number that gives the eigenvalues of the total spin operator, i.e.,  ${\bf S}^2=S(S+1)$. The reason for that choice is because at the beginning of the annealing process, when $s=0$, the ground state of $H_I$ has maximal total spin. Since the total spin is conserved during the annealing process, we can restrict our analysis to this subspace.

We use the total-spin basis
\begin{align}
  {\bf S}^2  |S,M \rangle &= S (S+1) |S,M \rangle\,,\\
S^z  |S,M \rangle &= M |S,M \rangle\,,  
\end{align}
and, following \cite{PhysRevA.95.042321}, we label the states as
\begin{equation}
    | w \rangle :=|S=N/2, M=N/2-w \rangle\,,
\end{equation}
where $w =0,1,2,...,N$. In this subspace, the matrix elements of the Hamiltonin $[H]_{\omega, \omega'}:=\langle \omega |H(s,\lambda)| \omega \rangle$ can be written as
\begin{align}
    [H]_{w,\, w}&= s \left[-\lambda N \left(1-\frac{2w}{N} \right)^p +(1-\lambda)\left(2w-\frac{2w^2}{N}+1 \right) \right]\,,\\
    [H]_{w+1,\,w}&= [H]_{w,\,w+1}=-(1-s)\sqrt{(N-w)(w+1)}\,,\\
    [H]_{w+2,\,w}&= [H]_{w,\,w+2}=\frac{1}{N} s(1-\lambda) \sqrt{(w+1)(w+2)(N-w)(N -w-1)}\,,
\end{align}
and all other elements are zero. 


Here, we study the so called $r$-parameter statistics, introduced in \cite{Oganesyan_2007}. Given a sorted spectrum  $\{E_n\}$ and the corresponding level spacing, $s_n=E_{n+1}-E_{n}$, the $r$-parameter is defined by
\begin{equation}
    r_n=\frac{\min\{s_n,s_{n-1}\}}{\max\{s_n,s_{n-1}\}}\,.
\end{equation}
This quantity is independent of the local density of states and so it does not require unfolding. We are interested in the average $r$-parameter, 
\begin{equation}
    \tilde{r} \equiv \frac{1}{N-2}\sum_{n=2}^{N-1} r_n \,,
\end{equation}
because it takes some simple benchmark values in the case of chaotic systems. For example, $\tilde{r}_\text{GOE} \approx 0.53590$, $\tilde{r}_\text{GUE} \approx 0.60266$, and $\tilde{r}_\text{GSE} \approx 0.67617$, for random matrices of the Gaussian Orthogonal Ensemble (GOE), Gaussian Unitary Ensemble (GUE) and Gaussian Simpletic Ensemble (GSE), respectively.\footnote{In the case of 3 $\times$ 3 random matrices, one can show that the distribution of consecutive level spacing ratios $P(r)$ is
$
    P(r)=\frac{1}{Z_{\alpha}}\frac{(r+r^2)^{\alpha}}{(1+r+r^2)^{1+\frac{3}{2}\alpha}}\,,
$
where $\alpha=1,2$ or $4$ for GOE, GUE, GSE respectively \cite{Atas_2013}. For a Poisson distribution, $P(r)=e^{-r}$. Here, $r$ denotes a continuous version of $r_n$. The average $r$-parameter is defined as
$    \tilde{r}= \int_{0}^{1} r\, P(r)\, dr\,$}. If the dynamics of a given systems is chaotic, one expects the $r$-parameter statistics to be described by random matrix theory, and $\tilde{r}$ to take one of the above-mentioned reference values.
For integrable systems, one generally expects the statistics of the ratio of consecutive level spacing to be described by a Poisson distribution. In this case,  $\tilde{r}_\text{Poisson}=0.38624$. Thus, the average $r$-parameter, $\tilde{r}$, diagnoses the chaotic behavior of the system, or the absence of it.

\begin{figure}[]
\centering
\begin{subfigure}[]{0.45\hsize}
    \includegraphics[width=\hsize]{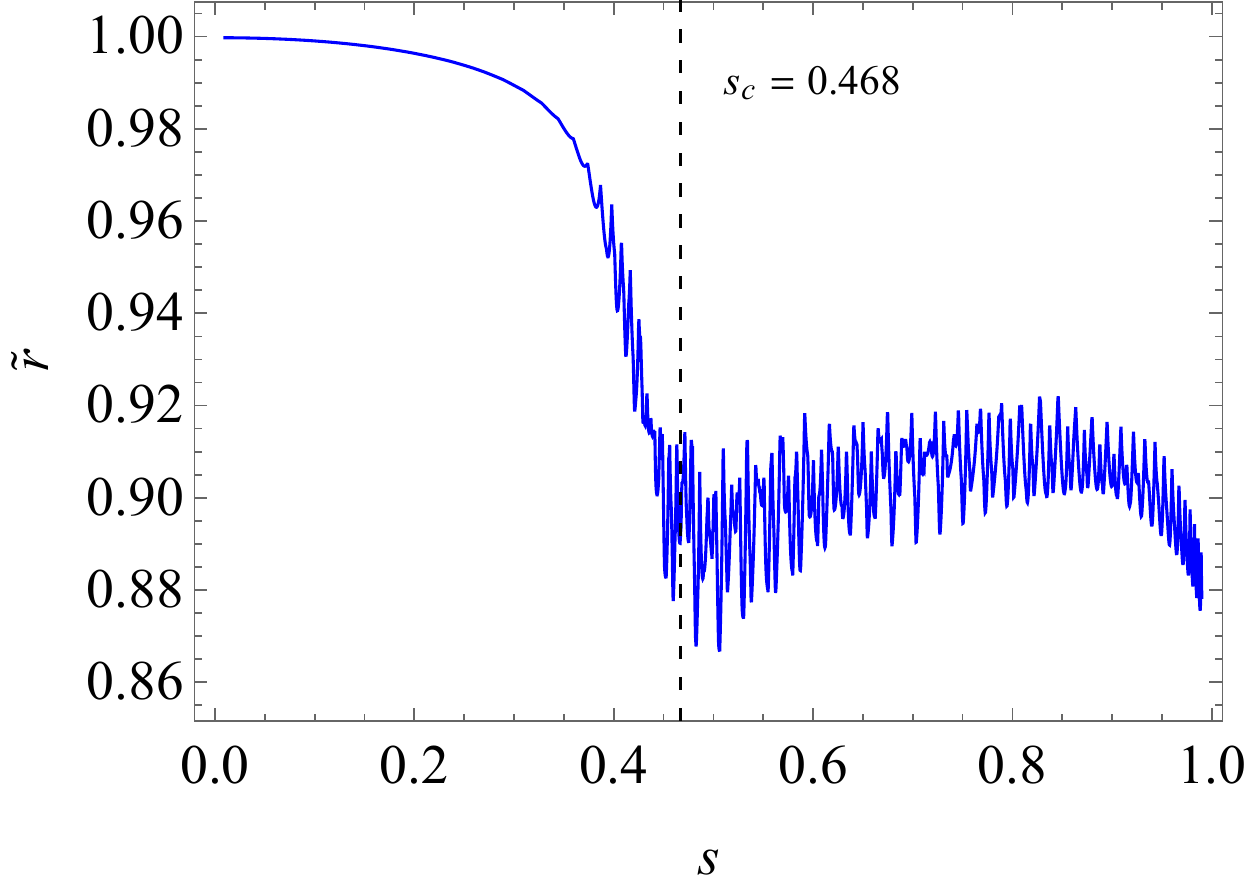}
    \caption{$p=5, \lambda=1.0$}
    \label{fig9:suba}
\end{subfigure}
\hspace{0.1cm}
\begin{subfigure}[]{0.45\hsize}
    \includegraphics[width=\hsize]{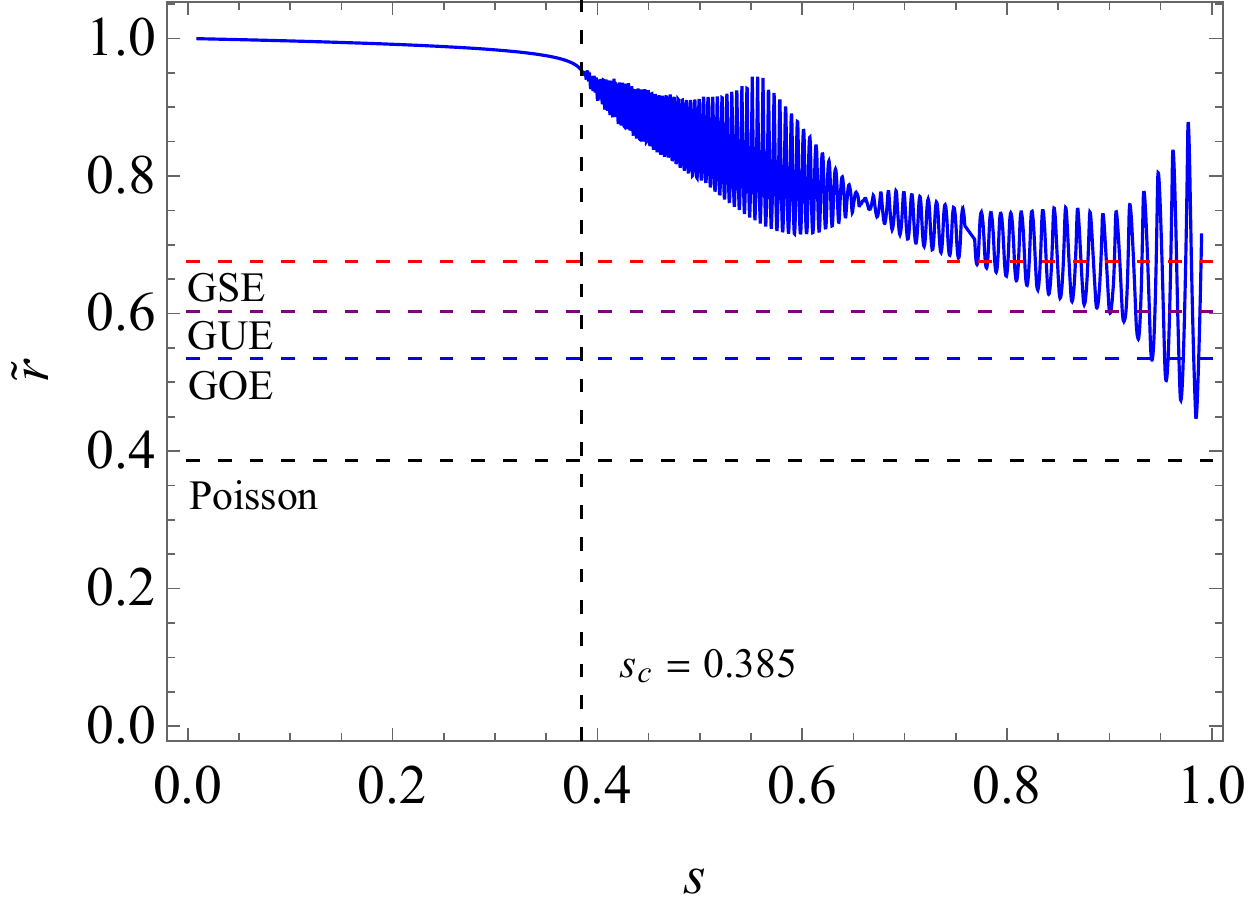}
    \caption{$p=5, \lambda=0.2$}
    \label{fig9:subb}
\end{subfigure}
\caption{ $s$ dependence of the average $r$-parameter ($\tilde{r}$) for the sector with maximal total spin. We set $N=100$. The red, purple, blue, and black horizontal lines correspond to the value of $\tilde{r}$ for GSE,  GUE, GOE, and Poisson distributions. The transition values of $s$ are indicated by a vertical black dashed line. These values agree with our OTOC computation, for example, Fig.~\ref{fig:Fbar}.  }
\label{fig:rparameter}
\end{figure}

Fig.~\ref{fig:rparameter} exhibits $\tilde{r}$ for the maximal spin sector of the Hamiltonain (\ref{eq-hamQA}). For small values of $s$, $\tilde{r}$ is close to one. As we increase $s$, the average $r$-parameter decreases slightly, and it starts to oscillate around the phase transition point.  For the anti-ferromagnetic interactions at $\lambda=0.2$ (Fig.~\ref{fig9:suba}) one can precisely find out the phase transition value of $s$, which agree with our OTOC result in Fig.~\ref{fig4:subb}. After the phase transition, in the F phase, $\tilde{r}$ oscillates wildly as we increase $s$, taking values that correspond to chaotic behavior for some specific $s$.\footnote{For some values of the annealing parameter, $\tilde{r}$ takes the reference values that are indicative of chaotic behavior, so one may think the system may be chaotic for these specific values of $s$. However, we find that the level spacing distributions at these values are not the ones of the random matrix distributions although they might have some vague resemblance.}
In the case with no anti-ferromagnetic interactions at $\lambda=1$ (Fig.~\ref{fig9:suba}), one can also estimate the phase transition value of $s$, which also agree with our OTOC result in Fig.~\ref{fig4:suba}, although in this case the phase transition point is less sharp. Moreover, in the F phase,  $\tilde{r}$ oscillates around some constant value which is far away from any reference value for $\tilde{r}$. Thus, in general, our system is not chaotic.

It is interesting to question whether the behavior of $\tilde{r}$ versus $s$ can indicate the order of the phase transition. Our result suggests that  it may be the case in the sense of the local average value of $\tilde{r}$ over $s$, $\tilde{r}_{\text{avg}}$, defined as 
\begin{equation} \label{eq-avgr}
\tilde{r}_{\text{avg}}(s)= \frac{1}{\Delta s} \int_{s-\Delta s/2}^{s+\Delta s/2} ds' \, \tilde{r}(s')\,,
\end{equation}
where $\Delta s$ is the range of the interval in $s$ that we use to compute the local average.
For the second order phase transition,{~Fig.~\ref{fig9:subb}}, the local average $\tilde{r}_{\text{avg}}(s)$ changes smoothly at the transition point, while for the first order transition, it changes abruptly~{Fig.~\ref{fig9:suba}}.

\section{Reverse Annealing and OTOCs}\label{sec:RA}
In this section, we study a relation between reverse annealing and OTOCs. Reverse annealing is a way to find a better classical solution than a given initial solution by starting from an appropriate classical state and gradually increasing and then decreasing the transverse magnetic field \cite{Perdomo-Ortiz2011}. The current D-Wave quantum annealer implements reverse annealing and the performance has been studied for various cases \cite{ikeda2019NSP,Venturelli2019,kind2018}. Some studies on the efficiency of reverse annealing are provided theoretically \cite{PhysRevA.98.022314} and numerically \cite{PhysRevA.100.052321}, where it is shown that, at least for the $p$-spin model, reverse annealing can turn first-order phase transition into second-order phase transition by choosing an appropriate process. 

The Hamiltonian of reverse annealing we will consider is 
\begin{equation}
    H(s,\lambda)=sH_T+(1-s)(1-\lambda)H_\text{init}+\Gamma(1-s)\lambda H_I\,,
\end{equation}
where $s,\lambda$ both take values in $[0,1]$ and $H_T$ is the Hamiltonian of the $p$-spin model \eqref{H000}.   $\Gamma$ tunes the strength of the transverse magnetic field $H_I$ \eqref{trans34}. $H_\text{init}$ is the initialization term
\begin{equation}
    H_\text{init}=-\sum_{i=1}^N\epsilon_i Z_i\,,
\end{equation}
which is $H(0,0)$ and $\epsilon_i\in\{-1,1\}$ is the $i$th component of the given classical initial state. Due to the high spatial symmetry of the $p$-spin model, without loss of generality we can assign $\epsilon_i$ as 
\begin{equation}
    \epsilon_i=
\begin{cases}
+1&\text{for~}i\le N-n\,,\\
-1&\text{for~}i> N-n\,,
\end{cases}
\end{equation}
where $n\in\{0,1,\cdots,N\}$. The system may be characterized by a parameter $c=1-\frac{n}{N}\in[0,1]$, which is the probability for $\epsilon_i$ to be $+1$. The initial magnetization is $2c-1$. A large $c$ implies the solution is close to the ground state of $H_T$.
According to \cite{PhysRevA.100.052321}, the $p$-spin model experiences phase transitions by reverse annealing. We will show that the phase transitions can be detected also by OTOCs. 

\begin{figure}[]
\centering
\begin{subfigure}[]{0.45\hsize}
    \includegraphics[width=\hsize]{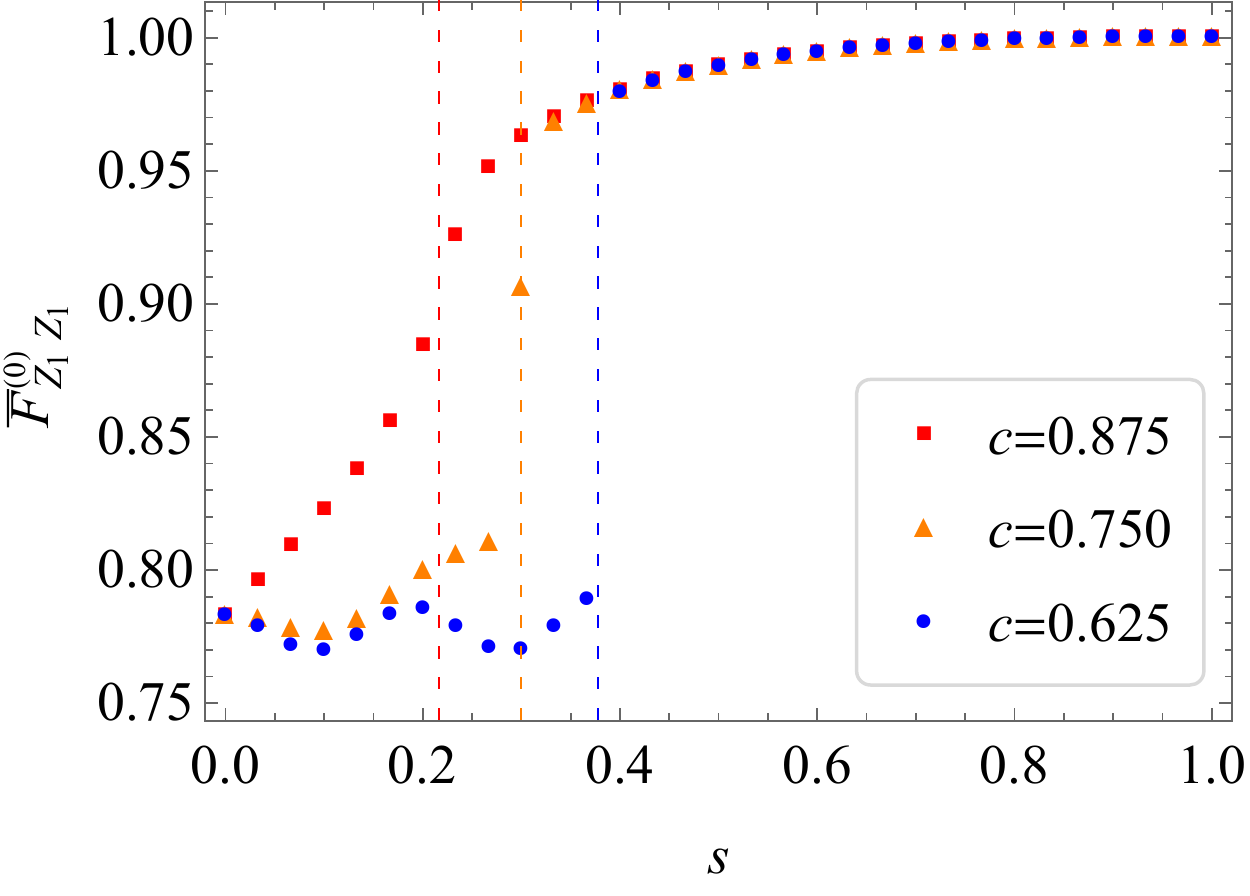}
    \caption{$p=5, \Gamma=1$}
\end{subfigure} \ \ \
\begin{subfigure}[]{0.45\hsize}
    \includegraphics[width=\hsize]{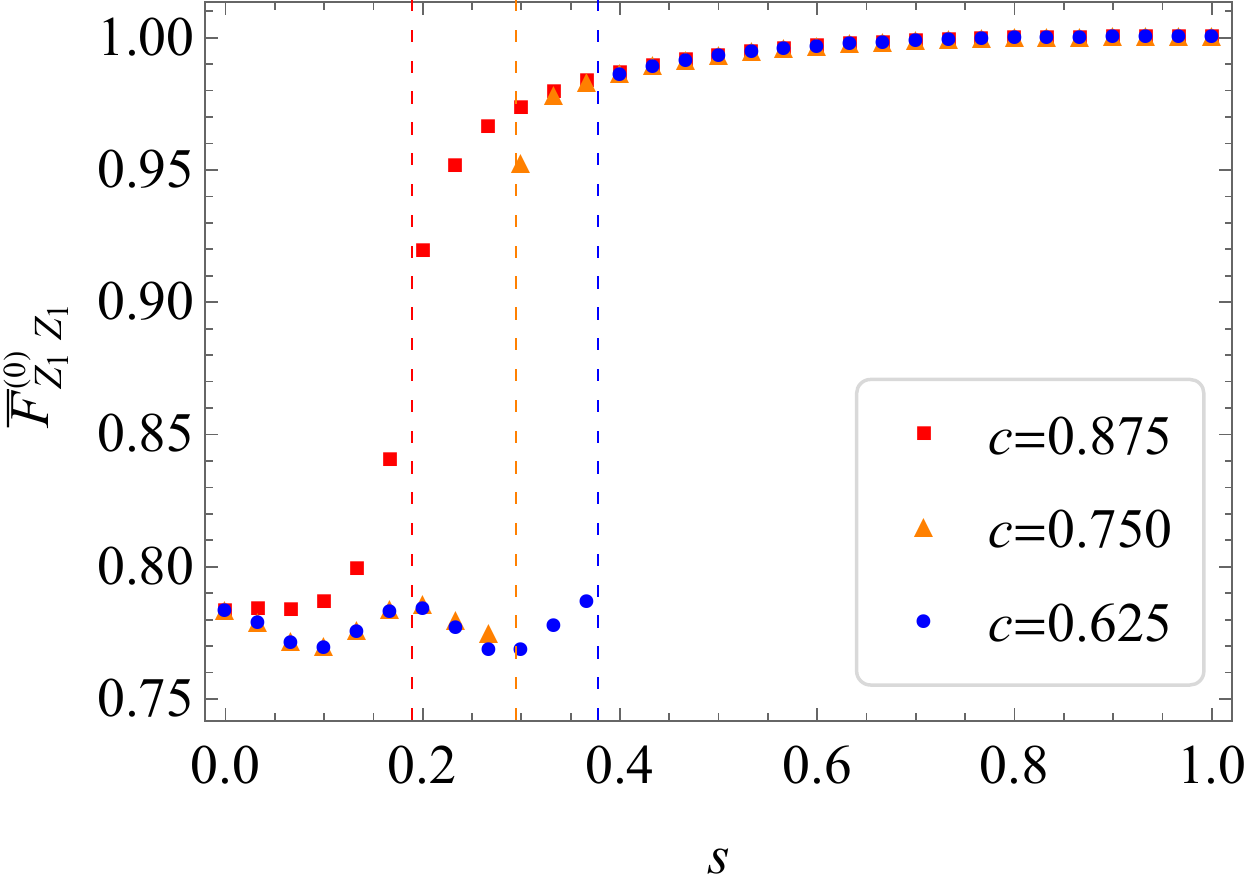}
    \caption{$p=11, \Gamma=1$}
\end{subfigure}
\par\bigskip
\begin{subfigure}[]{0.45\hsize}
    \includegraphics[width=\hsize]{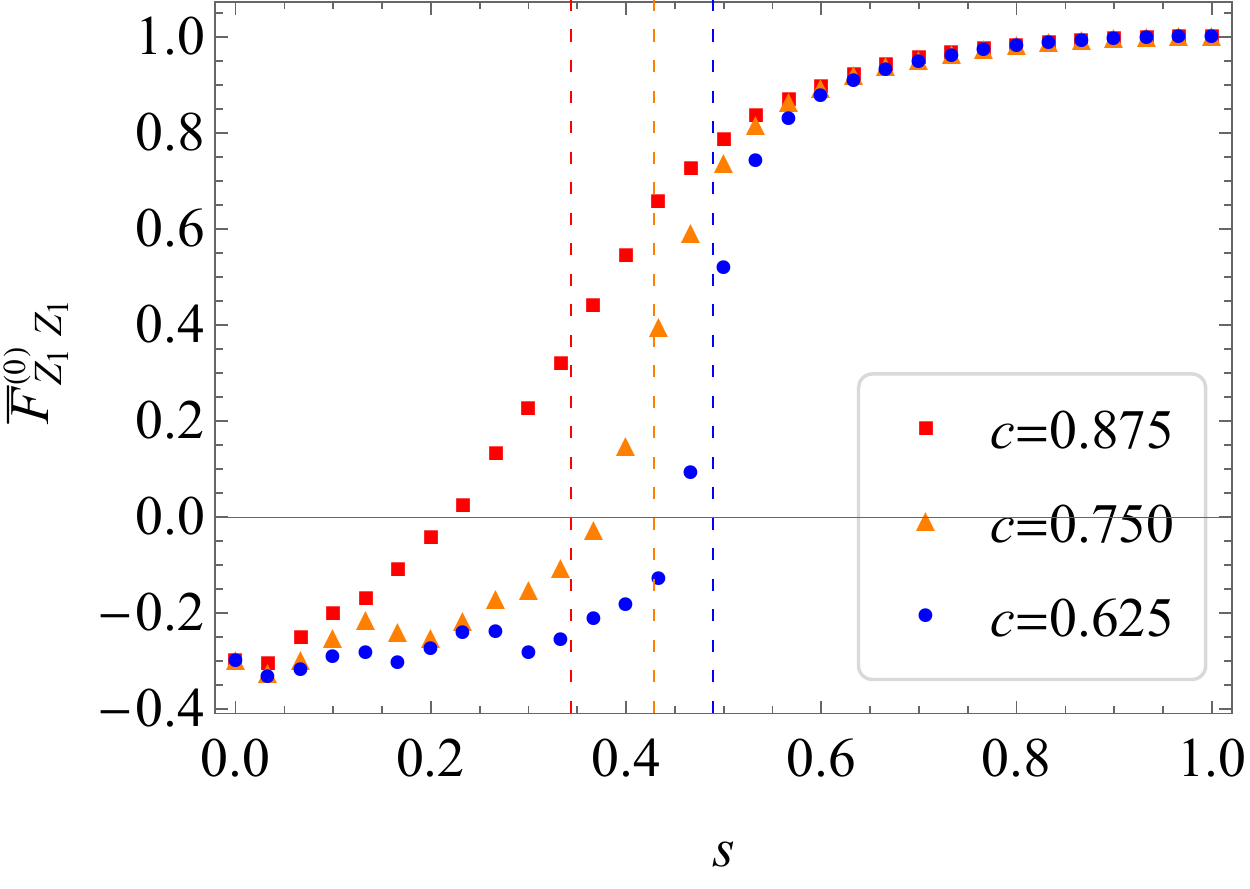}
    \caption{$p=5, \Gamma=5$}
\end{subfigure} \ \ \ 
\begin{subfigure}[]{0.45\hsize}
    \includegraphics[width=\hsize]{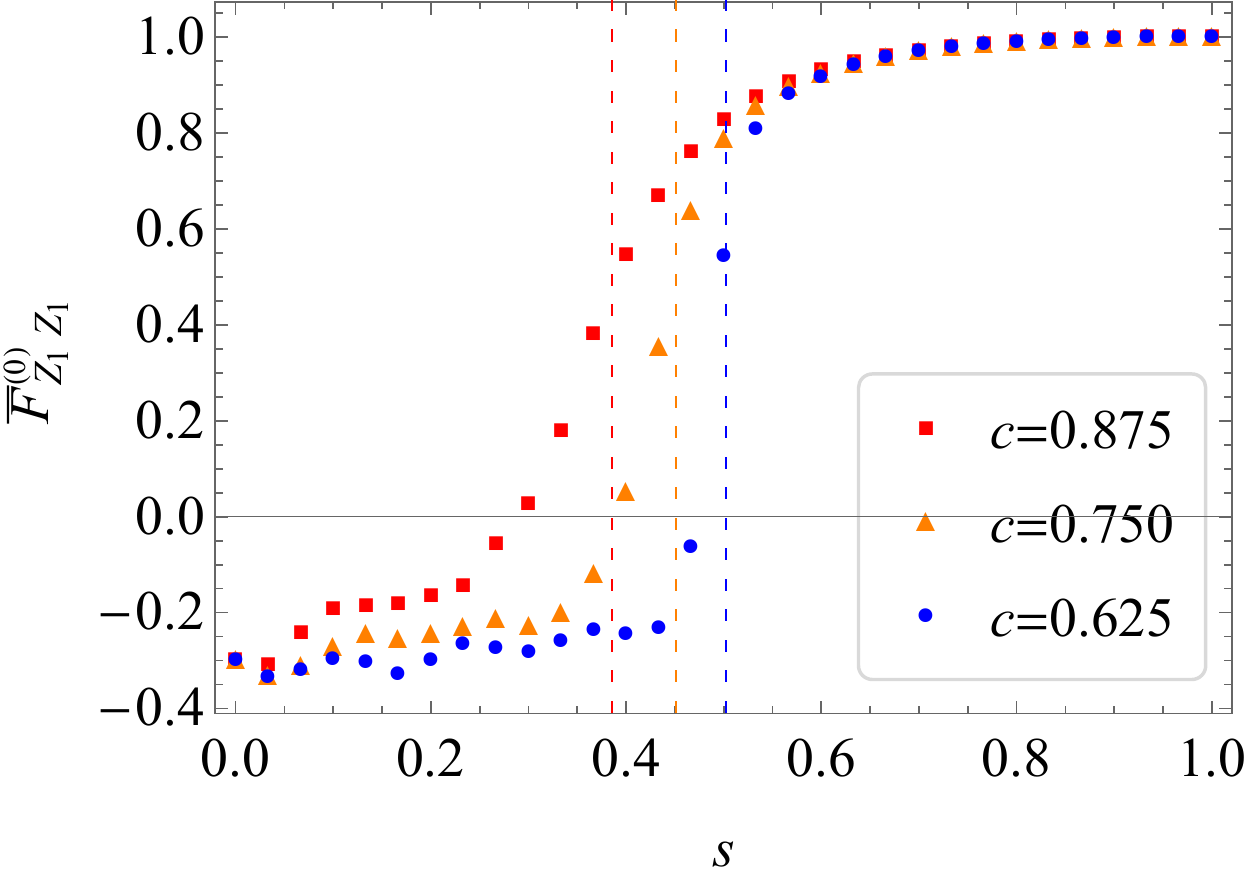}
    \caption{$p=11, \Gamma=5$}
\end{subfigure}
    \caption{$\bar{F}_{Z_i Z_i}^{(0)}$ versus $s$ during the reverse annealing process for $N=8$ and $\lambda=0.2$. The left panel show the results for $p=5$, while the right panel show the results for $p=11$. We set $\Gamma=1$ in the first row, and $\Gamma=5$ in the second row.}
    \label{fig:rev}
\end{figure}

In Fig.~\ref{fig:rev} we present the time averaged OTOCs of the form $\bar{F}_{Z_i Z_i}^{(0)}$ as an example. We have confirmed that other choices of operators show the same properties.
 $\bar{F}_{Z_i Z_i}^{(0)}$ grows around the dotted lines corresponding to the transition points obtained by computing the free energy following~\cite{PhysRevA.100.052321}. This confirms that OTOCs can diagnose the phase transition.
The smaller $c$ becomes, the larger the transition $s$ becomes. 
The bigger $\Gamma$ becomes, the bigger the transition points $s$ becomes. All these properties are also  consistent with \cite{PhysRevA.100.052321}. 
Furthermore, the curves for $c=0.625$ (blue circles) suggest a first order phase transition, while the curves for $c=0.875$ (red squares) suggest a second order phase transition. The results for $c=0.750$ depend on the value of $\Gamma$. For $\Gamma=1$, they suggest a first order phase transition, while for $\Gamma=5$ they suggest a second order phase transition. These are also consistent with \cite{PhysRevA.100.052321}, and provide more evidences that time-averaging OTOCs can diagnose also the order of phase transitions. The time-averaging OTOCs may play the role of an order parameter.

\section{Conclusions and Future Directions}\label{sec:fin}

In this work, we have studied quantum phase transitions  associated with quantum annealing and reverse quantum annealing for the ferromagnetic $p$-spin model from the point of view of quantum chaos. More specifically, we have shown that time-averaging OTOCs and the $r$-parameter statistics, which are usually used to diagnose quantum chaos, can also be used to diagnose quantum phase transitions.\footnote{Recently, quantum phase transitions from a QP phase into a F phase were observed in a different setup \cite{nie2019experimental}. Our results should be experimentally testable in the same manner.} 

In the case of quantum annealing (QA), the system's ground state is initially in a quantum paramagnetic (QP) phase ($s=0$). As we increase the value of $s$, the system displays a quantum phase transition to a ferromagnetic (F) phase. We observe that the time-averaging OTOC takes an approximately constant value which is different in each phase. The late-time average value of OTOCs of the form $F_{Z_iZ_j}^{(0)}=\langle Z_i(0) Z_j(t) Z_i(0) Z_j(t) \rangle$ vanishes in the QP phase, and it takes a non-zero value in the F phase. In contrast, in the case of OTOCs of the form $F_{X_iX_j}^{(0)}=\langle X_i(0) X_j(t) X_i(0) X_j(t) \rangle$, the late-time average value is non-zero in the QP phase and zero in the F phase. In both cases, despite having different values in each phase depending on the choice of operator in the OTOC, the time-averaging OTOC sharply diagnoses the quantum phase transitions. Time-averaging OTOCs made out of non-local operators, $\bar{F}^{(0)}_{S^z S^z}$ and $\bar{F}^{(0)}_{S^x S^x}$, show a qualitatively similar behavior to the local operators $\bar{F}^{(0)}_{Z_i Z_j}$ and $\bar{F}^{(0)}_{X_i X_j}$, respectively. 

The operator-dependence of $\bar{F}^{(0)}_{V W}$ occurs because we are considering ground state OTOCs.
The ground-state OTOCs is essentially controlled by the properties of the ground state and do not reflect any chaotic property of ferromagnetic $p$-spin model. To access the chaotic properties of the model, we need to consider thermal OTOCs at temperatures which are sufficiently larger than the ground state energy, in such a way that the OTOC receives contributions from excited states. In those cases, however, the thermal fluctuations washes out the special properties of the ground state, and the OTOCs are no longer useful in diagnosing phase transition. This can be seen in Fig.~\ref{fig:TOTOC}, which shows that the time-averaging thermal OTOCs vanish in both phases at high temperature.

The thermal OTOCs oscillate with time wildly around zero at late times, taking values which range roughly from -1 to 1. See Fig.~\ref{fig:OTOCthermal}.  We think this oscillatory behavior reflects two effects: finite-size effects and some sort of weak chaotic behavior. Finite-size effects account for the fact that OTOCs do not vanish at late-times in systems whose size is not very large. In fact, it was shown in \cite{Huang:2017fng} that for energy-conserving system the late-time value of OTOCs scale as an inverse polynomial in the system size. Since our calculation of thermal OTOCs were done for $N \sim 11$, we expect  finite-size effects in our results. The second effect, which we call {\it weak scrambling}, is based on the idea that, if the system is not strongly chaotic, or even if it is integrable, the thermal OTOC (for spatially separated operators) will not vanish at late times, but it will rather oscillate around zero, with the size of the oscillation being smaller for systems which are more chaotic. We give more evidence for these ideas in Appendix \ref{sec:app} using the chaotic Ising model. Similar ideas are also explored in \cite{Fortes_2019}.

From the above-mentioned results, we could confirm that time-averaging OTOCs provide useful order parameters to detect quantum phase transition, as suggested in \cite{Heyl_2018}. Interestingly, time-averaging OTOCs not only diagnose the phase transition, but they can also distinguish first-order and second order phase transitions. In the case of first-(second-) order phase transitions, the time averaging OTOCs display a discontinuous (continuous) behavior at the quantum transition point, just like ordinary order parameters. This suggests that time-averaging OTOCs might be useful in detecting phase transitions when the rigorous order parameter is unknown.  


The effectiveness of ground state OTOCs in detecting quantum phase transitions motivated us to question if we could also diagnose phase transitions using other probes of quantum chaos. In order to do that, we have studied how the average $r$-parameter behaves during the quantum annealing process. Fig.~\ref{fig:rparameter} shows that the average $r$-parameter clearly changes behavior around the quantum transition point, with the phase transition being more sharp in the case with anti-ferromagnetic interactions ($\lambda=0.2$). When $\lambda=1$, the average $r$-parameter, $\tilde{r}$, never takes any reference value characterizing random matrix behavior or integrability. In contrast, for $\lambda=0.2$, the curves of $\tilde{r}$ versus $s$ display large oscillations in the ferromagnetic phase, taking values that correspond to chaotic behavior at some specific $s$. However, we checked that the corresponding level spacing distributions are different from Poisson or random matrix distributions, although they might have some vague resemblance.
Interestingly, the local average of $\tilde{r}$ also can distinguish first and second order phase transition like OTOCs.
Since we have studied the $r$-parameter statistics in the maximal spin sector, and the ground state of the system belongs to this sector, it is natural that it can detect the ground state physics associated to the phase transition.

In the case of reverse QA, the $p$-model also displays quantum phase transitions, as shown in \cite{PhysRevA.100.052321}.  Similarly to the QA case, the time-averaging OTOC also can diagnose the quantum phase transition and its order. There is a quantitative difference from the QA case. In the beginning of the reverse annealing schedule, at $s=0$, the system already displays a non-zero value of $\bar{F}_{Z_iZ_i}^{(0)}$, which then increases as we increase the values of $s$. 

Finally, we make a comment about the presence of the anti-ferromagnetic interaction term, $H_{AF}$, which makes the Hamiltonian non-stoquastic and turns first-order phase transitions into second-order ones during the quantum annealing process. From the analysis of the level spacing statistic we observe that this term makes the dynamics richer in the ferromagnetic phase. In particular, for some values of $s$, the system displays some degree of level repulsion, which is typical of chaotic systems. This suggests a connection between chaos and the order of the phase transitions. In particular, it seems that the presence of chaos in the ground state sector may turn first order phase transitions into second order ones. It would be interesting to further investigate this possibility. 

\section*{Acknowledgement} We thank Koji Hahimoto, Norihiro Iizuka, Keisuke Fujii, Mikito Koshino, Kin-ya Oda, and Masaki Tezuka for comments and stimulating discussion. Our collaboration started when the authors attended at the 13th Kavli Asian Winter School on Strings, Particles and Cosmology, and developed while they were participating in Quantum Information and String Theory 2019 at YITP. We thank the conference organizers. A part of this work was completed when KI visited Brookhaven National Laboratory (BNL) and he thanks the members at RIKEN BNL Research Center (RBRC), especially Taku Izubuchi, Dmitri Kharzeev, Yuta Kikuchi, and Akio Tomiya for hospitality and useful discussion. He was also benefited from communication with Akinori Tanaka. KI was partly supported by Grant-in-Aid for JSPS Research Fellow, No. 19J11073. VJ, KK and KH were supported in part by
Basic Science Research Program through the National Research Foundation of Korea(NRF)
funded by the Ministry of Science, ICT \& Future Planning(NRF2017R1A2B4004810) and
GIST Research Institute(GRI) grant funded by the GIST in 2020.

\appendix

\section{Chaotic/Integrable Transition in the Ising Spin Chain}\label{sec:app}
In this section we review the time-behavior of OTOCs for the Ising model with transverse and magnetic fields,
\begin{equation} \label{eq-ham-Ising}
    H=-\sum_{i=1}^{L-1} Z_i Z_{i+1}-\sum_{i=1}^{L} \left(h_x X_i+h_z Z_i \right)\,, 
\end{equation}
where we chose open boundary conditions. The Hamiltonian (\ref{eq-ham-Ising}) is known to display chaotic behavior when $(h_x,h_z)=(-1.05,0.5)$ \cite{2011PhRvL.106e0405B}. The system is integrable if either $h_x$ or $h_z$ vanish.

We will be interested in the behavior of thermal OTOCs of the form
\begin{equation}
    F_{Z_i Z_j}(t)=\langle Z_i(0) Z_j(t) Z_i(0) Z_j(t)\rangle \,,
\end{equation}
where the expectation value is taken in a thermal state.
\begin{figure}[]
\centering
\begin{subfigure}[]{0.45\hsize} 
    \includegraphics[width=\hsize]{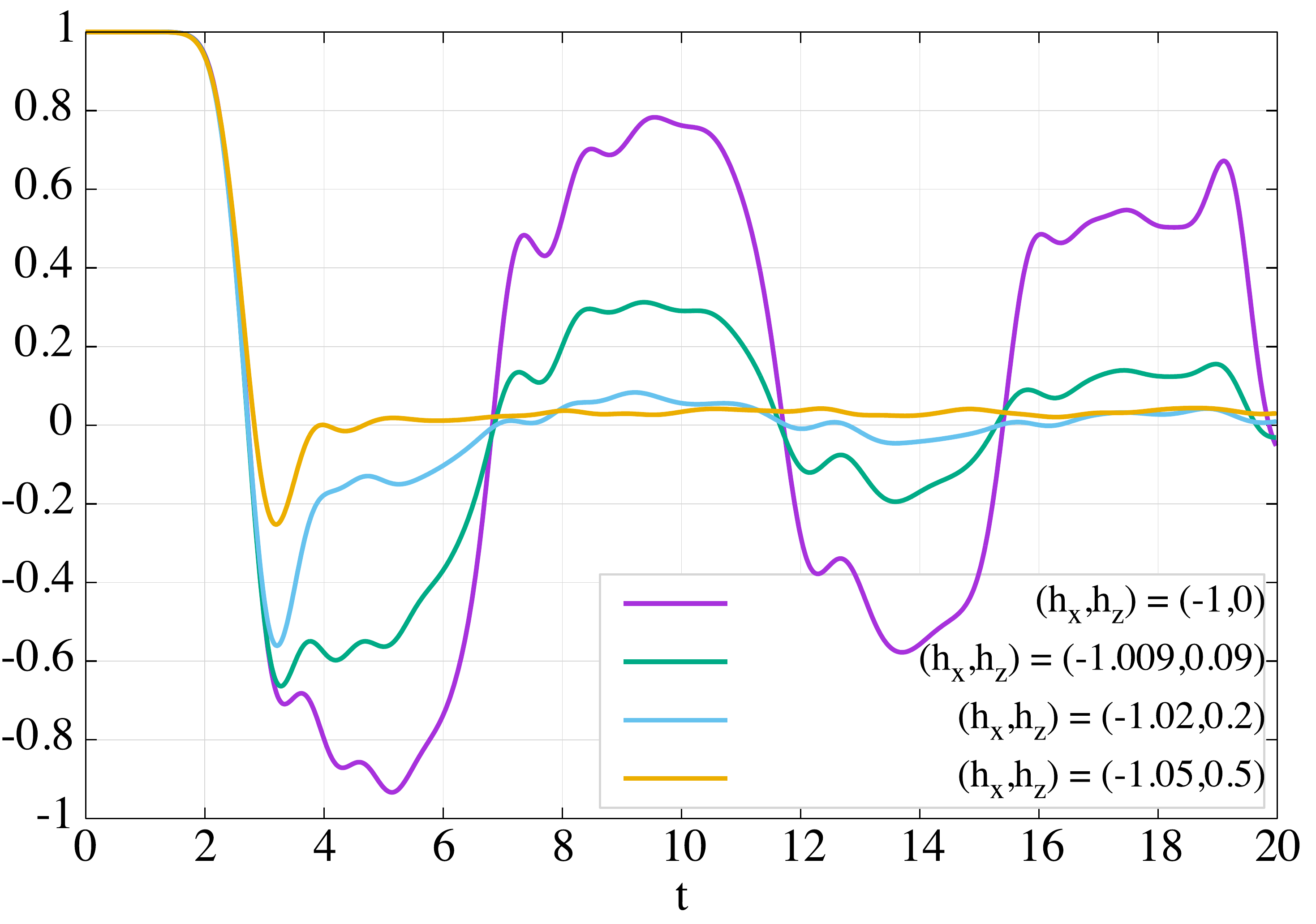}
   \caption{$\beta = 0$}
   \label{khbny1}
\end{subfigure} 
\hspace{0.5cm}
\begin{subfigure}[]{0.45\hsize} 
    \includegraphics[width=\hsize]{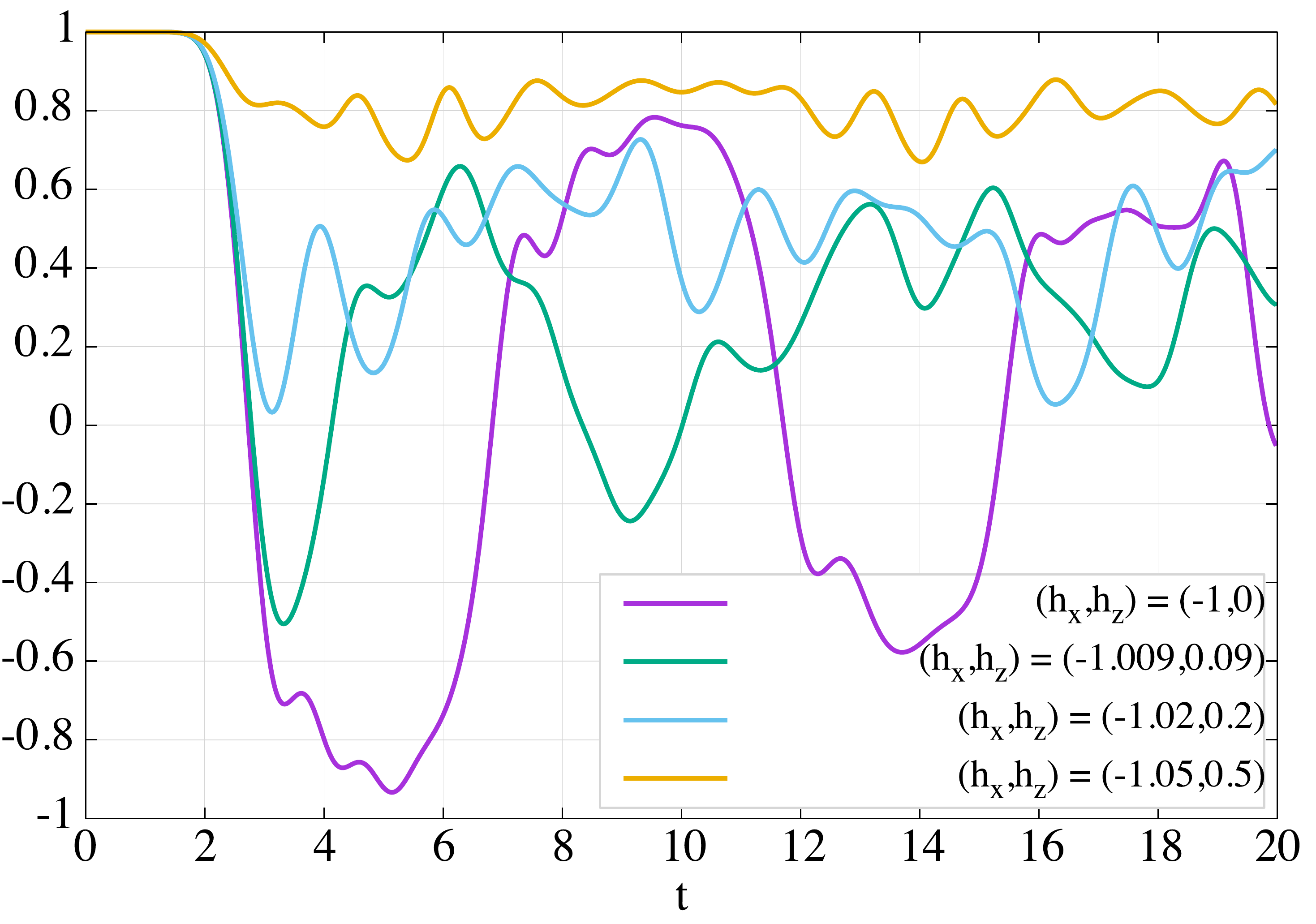}
    \caption{$\beta = 50$}
    \label{khbny2}
\end{subfigure} 
    \caption{Thermal OTOCs, $F_{Z_i Z_j}(t)=\langle Z_i(0) Z_j(t) Z_i(0) Z_j(t)\rangle$, for the Ising model with longitudinal and transverse magnetic fields ($L=8$). }
    \label{fig:otocs-Ising}
\end{figure}
In Fig.~\ref{fig:otocs-Ising} we study the behavior of OTOCs across the integrable/chaotic transition that takes place as we change the parameters from the integrable point $(h_x,h_z)=(-1,0)$ to the strongly chaotic point $(h_x,h_z)=(-1.05,0.5)$. 

Let us first discuss the infinite-temperature results, which are shown in Fig.~\ref{khbny1}.
At the strongly chaotic point $(h_x,h_z)=(-1.05,0.5)$, the OTOC vanishes at late times, which is a signal of scrambling \footnote{Here, by scrambling, we mean the late-time vanishing of OTOCs. In the literature, people sometimes use the term scrambling to refer to the exponential behavior of OTOCs.}. As we move away from the strongly chaotic point, the OTOC starts to oscillate around zero, and the amplitude of the oscillations increase as we approach the integrable point. We understand this behavior as a manifestation of {\it weak scrambling}, in which the OTOCs do not vanish at late times, but rather oscillates around zero, with the amplitude of the oscillations being smaller for more chaotic cases.

The behavior of the OTOCs change  as we reduce the system's temperature.  Fig.~\ref{khbny2} shows the results for $\beta=50$, which are almost same as the corresponding ground state OTOCs. 
Interestingly, the  OTOCs at the integrable point $(h_x,h_z)=(-1,0)$  do not seem to depend on the temperature. 
For all other values of $(h_x,h_z)$, the late-time value of the OTOC does not approach zero at late times. Instead, it oscillates around some constant value that depends on $(h_x,h_z)$. This shows that the OTOC no longer displays universal behavior at small temperatures, being actually controlled by the ground state physics. Note that this happens even at the strong chaotic point $(h_x,h_z)=(-1.05,0.5)$. We understand this phenomenon as an absence of scrambling at low temperatures.

The observed behavior of the OTOCs for the quantum annealing Hamiltonian (\ref{eq-hamQA}) are qualitatively similar to the ones obtained for the Ising model when we move slightly away from the integrable point, e.g. $(h_x,h_z)=(-1.02,.02)$. In both cases the thermal OTOCs (for sufficiently high temperatures) oscillate around zero taking some $\mathcal{O}(1)$ values. This suggests that the system described by the Hamiltonian (\ref{eq-hamQA}) is close to be integrable.

\bibliography{references}

\providecommand{\href}[2]{#2}\begingroup\raggedright\begin{thebibliography}{10}

\bibitem{larkin1969quasiclassical}
A.~Larkin and Y.~N. Ovchinnikov, \emph{Quasiclassical method in the theory of
  superconductivity}, {\emph{Sov Phys JETP} {\bfseries 28} (1969) 1200}.

\bibitem{Kitaev14}
A.~{Kitaev}, ``Hidden correlations in the hawking radiation and thermal
  noise.''

\bibitem{Maldacena:2015waa}
J.~Maldacena, S.~H. Shenker and D.~Stanford, \emph{{A bound on chaos}},
  \href{https://doi.org/10.1007/JHEP08(2016)106}{\emph{JHEP} {\bfseries 08}
  (2016) 106} [\href{https://arxiv.org/abs/1503.01409}{{\ttfamily
  1503.01409}}].

\bibitem{Jahnke:2018off}
V.~Jahnke, \emph{{Recent developments in the holographic description of quantum
  chaos}}, \href{https://doi.org/10.1155/2019/9632708}{\emph{Adv. High Energy
  Phys.} {\bfseries 2019} (2019) 9632708}
  [\href{https://arxiv.org/abs/1811.06949}{{\ttfamily 1811.06949}}].

\bibitem{2019PhRvX...9a1006Y}
B.~{Yoshida} and N.~Y. {Yao}, \emph{{Disentangling Scrambling and Decoherence
  via Quantum Teleportation}},
  \href{https://doi.org/10.1103/PhysRevX.9.011006}{\emph{Physical Review X}
  {\bfseries 9} (2019) 011006}
  [\href{https://arxiv.org/abs/1803.10772}{{\ttfamily 1803.10772}}].

\bibitem{2017NatPh..13..781G}
M.~{G{\"a}rttner}, J.~G. {Bohnet}, A.~{Safavi-Naini}, M.~L. {Wall}, J.~J.
  {Bollinger} and A.~M. {Rey}, \emph{{Measuring out-of-time-order correlations
  and multiple quantum spectra in a trapped-ion quantum magnet}},
  \href{https://doi.org/10.1038/nphys4119}{\emph{Nature Physics} {\bfseries 13}
  (2017) 781} [\href{https://arxiv.org/abs/1608.08938}{{\ttfamily
  1608.08938}}].

\bibitem{2017PhRvX...7c1011L}
J.~{Li}, R.~{Fan}, H.~{Wang}, B.~{Ye}, B.~{Zeng}, H.~{Zhai} et~al.,
  \emph{{Measuring Out-of-Time-Order Correlators on a Nuclear Magnetic
  Resonance Quantum Simulator}},
  \href{https://doi.org/10.1103/PhysRevX.7.031011}{\emph{Physical Review X}
  {\bfseries 7} (2017) 031011}
  [\href{https://arxiv.org/abs/1609.01246}{{\ttfamily 1609.01246}}].

\bibitem{2019NatCo..10.1581L}
R.~J. {Lewis-Swan}, A.~{Safavi-Naini}, J.~J. {Bollinger} and A.~M. {Rey},
  \emph{{Unifying scrambling, thermalization and entanglement through
  measurement of fidelity out-of-time-order correlators in the Dicke model}},
  \href{https://doi.org/10.1038/s41467-019-09436-y}{\emph{Nature
  Communications} {\bfseries 10} (2019) 1581}
  [\href{https://arxiv.org/abs/1808.07134}{{\ttfamily 1808.07134}}].

\bibitem{BGS}
O.~Bohigas, M.~J. Giannoni and C.~Schmit, \emph{Characterization of chaotic
  quantum spectra and universality of level fluctuation laws},
  \href{https://doi.org/10.1103/PhysRevLett.52.1}{\emph{Phys. Rev. Lett.}
  {\bfseries 52} (1984) 1}.

\bibitem{BTZ-Poisson}
B.~M. Victor, T.~M. and Z.~J. Michael, \emph{Level clustering in the regular
  spectrum},
  \href{https://doi.org/http://doi.org/10.1098/rspa.1977.0140}{\emph{Proc. R.
  Soc. Lond. A} {\bfseries 356} (1977) }.

\bibitem{Berry-80}
M.~V. Berry, \emph{Quantizing a classically ergodic system: Sinai's billiard
  and the kkr method},
  \href{https://doi.org/https://doi.org/10.1016/0003-4916(81)90189-5}{\emph{Annals
  of Physics} {\bfseries 131} (1981) 163}.

\bibitem{Dag:2019yqu}
C.~B. Dağ, K.~Sun and L.~M. Duan, \emph{{Detection of Quantum Phases via
  Out-of-Time-Order Correlators}},
  \href{https://doi.org/10.1103/PhysRevLett.123.140602}{\emph{Phys. Rev. Lett.}
  {\bfseries 123} (2019) 140602}
  [\href{https://arxiv.org/abs/1902.05041}{{\ttfamily 1902.05041}}].

\bibitem{doi:10.1002/andp.201900270}
Z.-H. Sun, J.-Q. Cai, Q.-C. Tang, Y.~Hu and H.~Fan, \emph{Out-of-time-order
  correlators and quantum phase transitions in the rabi and dicke models},
  \href{https://doi.org/10.1002/andp.201900270}{\emph{Annalen der Physik}
  {\bfseries n/a} 1900270}
  [\href{https://arxiv.org/abs/https://onlinelibrary.wiley.com/doi/pdf/10.1002/andp.201900270}{{\ttfamily
  https://onlinelibrary.wiley.com/doi/pdf/10.1002/andp.201900270}}].

\bibitem{2018arXiv181201920W}
Q.~{Wang} and F.~{P{\'e}rez-Bernal}, \emph{{Probing excited-state quantum phase
  transition in a quantum many body system via out-of-time-ordered
  correlator}}, {\emph{arXiv e-prints} (2018) arXiv:1812.01920}
  [\href{https://arxiv.org/abs/1812.01920}{{\ttfamily 1812.01920}}].

\bibitem{Garcia-Garcia:2019poj}
A.~M. Garc\'\i{}a-Garc\'\i{}a, T.~Nosaka, D.~Rosa and J.~J. Verbaarschot,
  \emph{{Quantum chaos transition in a two-site Sachdev-Ye-Kitaev model dual to
  an eternal traversable wormhole}},
  \href{https://doi.org/10.1103/PhysRevD.100.026002}{\emph{Phys. Rev. D}
  {\bfseries 100} (2019) 026002}
  [\href{https://arxiv.org/abs/1901.06031}{{\ttfamily 1901.06031}}].

\bibitem{nosaka2020quantum}
T.~Nosaka and T.~Numasawa, \emph{Quantum chaos, thermodynamics and black hole
  microstates in the mass deformed syk model},  2020.

\bibitem{nosaka2020chaos}
T.~Nosaka and T.~Numasawa, \emph{Chaos exponents of syk traversable wormholes},
   2020.

\bibitem{Vojta_2003}
M.~Vojta, \emph{Quantum phase transitions},
  \href{https://doi.org/10.1088/0034-4885/66/12/r01}{\emph{Reports on Progress
  in Physics} {\bfseries 66} (2003) 2069–2110}.

\bibitem{PhysRevE.58.5355}
T.~Kadowaki and H.~Nishimori, \emph{Quantum annealing in the transverse ising
  model}, \href{https://doi.org/10.1103/PhysRevE.58.5355}{\emph{Phys. Rev. E}
  {\bfseries 58} (1998) 5355}.

\bibitem{2000quant.ph..1106F}
E.~{Farhi}, J.~{Goldstone}, S.~{Gutmann} and M.~{Sipser}, \emph{{Quantum
  Computation by Adiabatic Evolution}}, {\emph{ArXiv e-prints} (2000) quant}
  [\href{https://arxiv.org/abs/quant-ph/0001106}{{\ttfamily
  quant-ph/0001106}}].

\bibitem{Damski_2013}
B.~Damski and M.~M. Rams, \emph{Exact results for fidelity susceptibility of
  the quantum ising model: the interplay between parity, system size, and
  magnetic field},
  \href{https://doi.org/10.1088/1751-8113/47/2/025303}{\emph{Journal of Physics
  A: Mathematical and Theoretical} {\bfseries 47} (2013) 025303}.

\bibitem{PhysRevB.71.224420}
S.~Dusuel and J.~Vidal, \emph{Continuous unitary transformations and
  finite-size scaling exponents in the lipkin-meshkov-glick model},
  \href{https://doi.org/10.1103/PhysRevB.71.224420}{\emph{Phys. Rev. B}
  {\bfseries 71} (2005) 224420}.

\bibitem{PFEUTY197079}
P.~Pfeuty, \emph{The one-dimensional ising model with a transverse field},
  {\emph{Annals of Physics} {\bfseries 57} (1970) 79 }.

\bibitem{PhysRevLett.109.030502}
C.~R. Laumann, R.~Moessner, A.~Scardicchio and S.~L. Sondhi, \emph{Quantum
  adiabatic algorithm and scaling of gaps at first-order quantum phase
  transitions},
  \href{https://doi.org/10.1103/PhysRevLett.109.030502}{\emph{Phys. Rev. Lett.}
  {\bfseries 109} (2012) 030502}.

\bibitem{doi:10.7566/JPSJ.82.114004}
J.~Tsuda, Y.~Yamanaka and H.~Nishimori, \emph{Energy gap at first-order quantum
  phase transitions: An anomalous case},
  \href{https://doi.org/10.7566/JPSJ.82.114004}{\emph{Journal of the Physical
  Society of Japan} {\bfseries 82} (2013) 114004}
  [\href{https://arxiv.org/abs/https://doi.org/10.7566/JPSJ.82.114004}{{\ttfamily
  https://doi.org/10.7566/JPSJ.82.114004}}].

\bibitem{Perdomo-Ortiz2011}
A.~Perdomo-Ortiz, S.~E. Venegas-Andraca and A.~Aspuru-Guzik, \emph{A study of
  heuristic guesses for adiabatic quantum computation}, {\emph{Quantum
  Information Processing} {\bfseries 10} (2011) 33}.

\bibitem{PhysRevA.95.042321}
Y.~Susa, J.~F. Jadebeck and H.~Nishimori, \emph{Relation between quantum
  fluctuations and the performance enhancement of quantum annealing in a
  nonstoquastic hamiltonian},
  \href{https://doi.org/10.1103/PhysRevA.95.042321}{\emph{Phys. Rev. A}
  {\bfseries 95} (2017) 042321}.

\bibitem{2019arXiv191107186P}
G.~{Passarelli}, K.-W. {Yip}, D.~A. {Lidar}, H.~{Nishimori} and P.~{Lucignano},
  \emph{{Reverse quantum annealing of the $p$-spin model with relaxation}},
  {\emph{arXiv e-prints} (2019) arXiv:1911.07186}
  [\href{https://arxiv.org/abs/1911.07186}{{\ttfamily 1911.07186}}].

\bibitem{2011PhRvA..83b2327F}
M.~{Filippone}, S.~{Dusuel} and J.~{Vidal}, \emph{{Quantum phase transitions in
  fully connected spin models: An entanglement perspective}},
  \href{https://doi.org/10.1103/PhysRevA.83.022327}{\emph{Phys. Rev. A}
  {\bfseries 83} (2011) 022327}
  [\href{https://arxiv.org/abs/1101.3654}{{\ttfamily 1101.3654}}].

\bibitem{Heyl_2018}
M.~Heyl, F.~Pollmann and B.~Dóra, \emph{Detecting equilibrium and dynamical
  quantum phase transitions in ising chains via out-of-time-ordered
  correlators},
  \href{https://doi.org/10.1103/physrevlett.121.016801}{\emph{Physical Review
  Letters} {\bfseries 121} (2018) }.

\bibitem{2017PhRvB..96e4503S}
H.~{Shen}, P.~{Zhang}, R.~{Fan} and H.~{Zhai}, \emph{{Out-of-time-order
  correlation at a quantum phase transition}},
  \href{https://doi.org/10.1103/PhysRevB.96.054503}{\emph{Phys. Rev. B}
  {\bfseries 96} (2017) 054503}
  [\href{https://arxiv.org/abs/1608.02438}{{\ttfamily 1608.02438}}].

\bibitem{PhysRevLett.123.140602}
C.~B. Da\ifmmode~\breve{g}\else \u{g}\fi{}, K.~Sun and L.-M. Duan,
  \emph{Detection of quantum phases via out-of-time-order correlators},
  \href{https://doi.org/10.1103/PhysRevLett.123.140602}{\emph{Phys. Rev. Lett.}
  {\bfseries 123} (2019) 140602}.

\bibitem{shukla2020outoftimeorder}
R.~K. Shukla, G.~K. Naik and S.~K. Mishra, \emph{Out-of-time-order correlation
  and detection of phase structure in floquet transverse ising spin system},
  2020.

\bibitem{2006quant.ph..6140B}
S.~{Bravyi}, D.~P. {DiVincenzo}, R.~I. {Oliveira} and B.~M. {Terhal},
  \emph{{The Complexity of Stoquastic Local Hamiltonian Problems}},
  {\emph{arXiv e-prints} (2006) quant}
  [\href{https://arxiv.org/abs/quant-ph/0606140}{{\ttfamily
  quant-ph/0606140}}].

\bibitem{2012PhRvE..85e1112S}
Y.~{Seki} and H.~{Nishimori}, \emph{{Quantum annealing with antiferromagnetic
  fluctuations}}, \href{https://doi.org/10.1103/PhysRevE.85.051112}{\emph{Phys.
  Rev. E} {\bfseries 85} (2012) 051112}
  [\href{https://arxiv.org/abs/1203.2418}{{\ttfamily 1203.2418}}].

\bibitem{J_rg_2010}
T.~J{\"o}rg, F.~Krzakala, J.~Kurchan, A.~C. Maggs and J.~Pujos, \emph{Energy
  gaps in quantum first-order mean-field{\textendash}like transitions: The
  problems that quantum annealing cannot solve}, {\emph{{EPL} (Europhysics
  Letters)} {\bfseries 89} (2010) 40004}.

\bibitem{ikeda_universal_2020}
K.~Ikeda, \emph{Universal computation with quantum fields},
  \href{https://doi.org/10.1007/s11128-020-02811-5}{\emph{Quantum Information
  Processing} {\bfseries 19} (2020) 331}.

\bibitem{Ahn:2019rnq}
Y.~Ahn, V.~Jahnke, H.-S. Jeong and K.-Y. Kim, \emph{{Scrambling in Hyperbolic
  Black Holes: shock waves and pole-skipping}},
  \href{https://doi.org/10.1007/JHEP10(2019)257}{\emph{JHEP} {\bfseries 10}
  (2019) 257} [\href{https://arxiv.org/abs/1907.08030}{{\ttfamily
  1907.08030}}].

\bibitem{Jahnke:2019gxr}
V.~Jahnke, K.-Y. Kim and J.~Yoon, \emph{{On the Chaos Bound in Rotating Black
  Holes}}, \href{https://doi.org/10.1007/JHEP05(2019)037}{\emph{JHEP}
  {\bfseries 05} (2019) 037}
  [\href{https://arxiv.org/abs/1903.09086}{{\ttfamily 1903.09086}}].

\bibitem{Huang:2017fng}
Y.~Huang, F.~G. S.~L. Brandão and Y.-L. Zhang, \emph{{Finite-size scaling of
  out-of-time-ordered correlators at late times}},
  \href{https://doi.org/10.1103/PhysRevLett.123.010601}{\emph{Phys. Rev. Lett.}
  {\bfseries 123} (2019) 010601}
  [\href{https://arxiv.org/abs/1705.07597}{{\ttfamily 1705.07597}}].

\bibitem{Khemani:2018sdn}
V.~Khemani, D.~A. Huse and A.~Nahum, \emph{{Velocity-dependent Lyapunov
  exponents in many-body quantum, semiclassical, and classical chaos}},
  \href{https://doi.org/10.1103/PhysRevB.98.144304}{\emph{Phys. Rev.}
  {\bfseries B98} (2018) 144304}
  [\href{https://arxiv.org/abs/1803.05902}{{\ttfamily 1803.05902}}].

\bibitem{Craps:2019rbj}
B.~Craps, M.~D. Clerck, D.~Janssens, V.~Luyten and C.~Rabideau, \emph{{Lyapunov
  growth in quantum spin chains}},
  \href{https://arxiv.org/abs/1908.08059}{{\ttfamily 1908.08059}}.

\bibitem{Gharibyan:2018fax}
H.~Gharibyan, M.~Hanada, B.~Swingle and M.~Tezuka, \emph{{Quantum Lyapunov
  Spectrum}}, \href{https://doi.org/10.1007/JHEP04(2019)082}{\emph{JHEP}
  {\bfseries 04} (2019) 082}
  [\href{https://arxiv.org/abs/1809.01671}{{\ttfamily 1809.01671}}].

\bibitem{Oganesyan_2007}
V.~Oganesyan and D.~A. Huse, \emph{Localization of interacting fermions at high
  temperature},
  \href{https://doi.org/10.1103/physrevb.75.155111}{\emph{Physical Review B}
  {\bfseries 75} (2007) }.

\bibitem{Atas_2013}
Y.~Y. Atas, E.~Bogomolny, O.~Giraud and G.~Roux, \emph{Distribution of the
  ratio of consecutive level spacings in random matrix ensembles},
  \href{https://doi.org/10.1103/physrevlett.110.084101}{\emph{Physical Review
  Letters} {\bfseries 110} (2013) }.

\bibitem{ikeda2019NSP}
K.~Ikeda, Y.~Nakamura and T.~S. Humble, \emph{Application of quantum annealing
  to nurse scheduling problem},
  \href{https://doi.org/10.1038/s41598-019-49172-3}{\emph{Scientific Reports}
  {\bfseries 9} (2019) 12837}.

\bibitem{Venturelli2019}
D.~Venturelli and A.~Kondratyev, \emph{Reverse quantum annealing approach to
  portfolio optimization problems}, {\emph{Quantum Machine Intelligence}
  {\bfseries 1} (2019) 17}.

\bibitem{kind2018}
A.~D. King, J.~Carrasquilla, J.~Raymond, I.~Ozfidan, E.~Andriyash, A.~Berkley
  et~al., \emph{Observation of topological phenomena in a programmable lattice
  of 1,800 qubits},
  \href{https://doi.org/10.1038/s41586-018-0410-x}{\emph{Nature} {\bfseries
  560} (2018) 456}.

\bibitem{PhysRevA.98.022314}
M.~Ohkuwa, H.~Nishimori and D.~A. Lidar, \emph{Reverse annealing for the fully
  connected $p$-spin model},
  \href{https://doi.org/10.1103/PhysRevA.98.022314}{\emph{Phys. Rev. A}
  {\bfseries 98} (2018) 022314}.

\bibitem{PhysRevA.100.052321}
Y.~Yamashiro, M.~Ohkuwa, H.~Nishimori and D.~A. Lidar, \emph{Dynamics of
  reverse annealing for the fully connected $p$-spin model},
  \href{https://doi.org/10.1103/PhysRevA.100.052321}{\emph{Phys. Rev. A}
  {\bfseries 100} (2019) 052321}.

\bibitem{nie2019experimental}
X.~Nie, B.-B. Wei, X.~Chen, Z.~Zhang, X.~Zhao, C.~Qiu et~al.,
  \emph{Experimental observation of equilibrium and dynamical quantum phase
  transitions via out-of-time-ordered correlators},  2019.

\bibitem{Fortes_2019}
E.~M. Fortes, I.~García-Mata, R.~A. Jalabert and D.~A. Wisniacki,
  \emph{Gauging classical and quantum integrability through out-of-time-ordered
  correlators},
  \href{https://doi.org/10.1103/physreve.100.042201}{\emph{Physical Review E}
  {\bfseries 100} (2019) }.

\bibitem{2011PhRvL.106e0405B}
M.~C. {Ba{\~n}uls}, J.~I. {Cirac} and M.~B. {Hastings}, \emph{{Strong and Weak
  Thermalization of Infinite Nonintegrable Quantum Systems}},
  \href{https://doi.org/10.1103/PhysRevLett.106.050405}{\emph{Phys. Rev. Lett.}
  {\bfseries 106} (2011) 050405}
  [\href{https://arxiv.org/abs/1007.3957}{{\ttfamily 1007.3957}}].

\end{thebibliography}\endgroup
\bibliographystyle{JHEP}

\end{document}